\documentclass[epj]{svjour}

\usepackage{graphics,epsf,graphicx,amssymb,subfig,amsmath,color}

\begin{document}
\title{A new deterministic model for chaotic reversals }
\author{Christophe Gissinger\inst{1} 
}
\institute{1 Department of Astrophysical
    Sciences/Princeton Plasma Physics Lab, 5 Peyton Hall, Princeton University,
    Princeton NJ USA.  19104-2688}
\date{Received: date / Revised version: date}
%
\abstract{We present a new chaotic system of three coupled ordinary
  differential equations, limited to quadratic nonlinear terms. A wide
  variety of dynamical regimes are reported. For some parameters,
  chaotic reversals of the amplitudes are produced by crisis-induced
  intermittency, following a mechanism different from what is
  generally observed in similar deterministic models. Despite its
  simplicity, this system therefore generates a rich dynamics, able to
  model more complex physical systems. In particular, a comparison
  with reversals of the magnetic field of the Earth shows a
  surprisingly good agreement, and highlights the relevance of
  deterministic chaos to describe geomagnetic field dynamics.  }
\maketitle

\section{Introduction}
\label{intro}
Deterministic chaotic models have been largely studied in the past few
decades, both on the numerical and theoretical sides
\cite{Wiggins90},\cite{Ott93},\cite{GuckHolmes}. In addition to their
interest to non-linear mathematics, these models often play an
important role in the comprehension of physical systems. Indeed, in
many situations, they offer a simple modelisation of experimental and
natural objects exhibiting complex dynamics. Particular interest has
been devoted to models involving only three equations, the minimum
number required to generate chaos in a continuous dynamical system
with time independent coefficients. Similarly, ordinary differential
equations of these types of models are generally restricted to the
lowest order terms. This approach ensures that numerical integration
yields chaotic dynamics relevant to describe the behaviors of complex
systems, meanwhile a theoretical study of the model is still
possible.\\

In this article, we focus on the dynamics of reversals, i.e. the
periodic or chaotic changes of sign of the amplitudes in dynamical
systems. Reversals between two symmetric states are observed in
several natural or experimental systems, among which we can mention
thermal convection, magnetohydrodynamics dynamos, or turbulent
atmospheric flows. Although these systems generally involve turbulent
states and a very large number of degrees of freedom, several sets of
deterministic equations have been proposed to model reversals, the
Lorenz model providing the most famous example \cite{Lorenz63}. As
typical examples of chaotic three-dimensional systems exhibiting
reversals, we can also mention the Rossler attractor \cite{Rossler76},
the Rikitake equations \cite{Rikitake58} and the Sprott models
\cite{Sprott94}. Such deterministic three-mode equations often provide
relevant toy models, but generally fail to describe the details of the
dynamics of the modeled system. In the case of reversals occurring in
fluids mechanics, in which turbulent fluctuations are important,
explanations in terms of stochastic processes are thus usually
preferred to nonlinear chaotic dynamics. Here, we propose a new model
for reversals, given by the following system of three coupled ordinary
differential equations:
\begin{eqnarray}
\label{mod_ampD}
\dot{Q} &=& \mu Q - VD \\
\label{mod_ampQ}
\dot{D} &=& -\nu D + VQ \\
\label{mod_ampV}
\dot{V} &=& \Gamma -V + QD
\end{eqnarray}
\noindent 
We will see that the type of dynamics generated by these equations is
relatively different from previous nonlinear chaotic models, and is
sometimes more similar to stochastic systems. It has been recently
proposed as a model for reversals of a turbulent magnetic dynamo
obtained in numerical simulations \cite{Gissinger10}. In the last
section of this article, we will discuss a physical interpretation of
this model in the framework of reversals of the Earth magnetic field.

\begin{figure*}[ht]
  \centerline{
             \includegraphics[height=62mm]{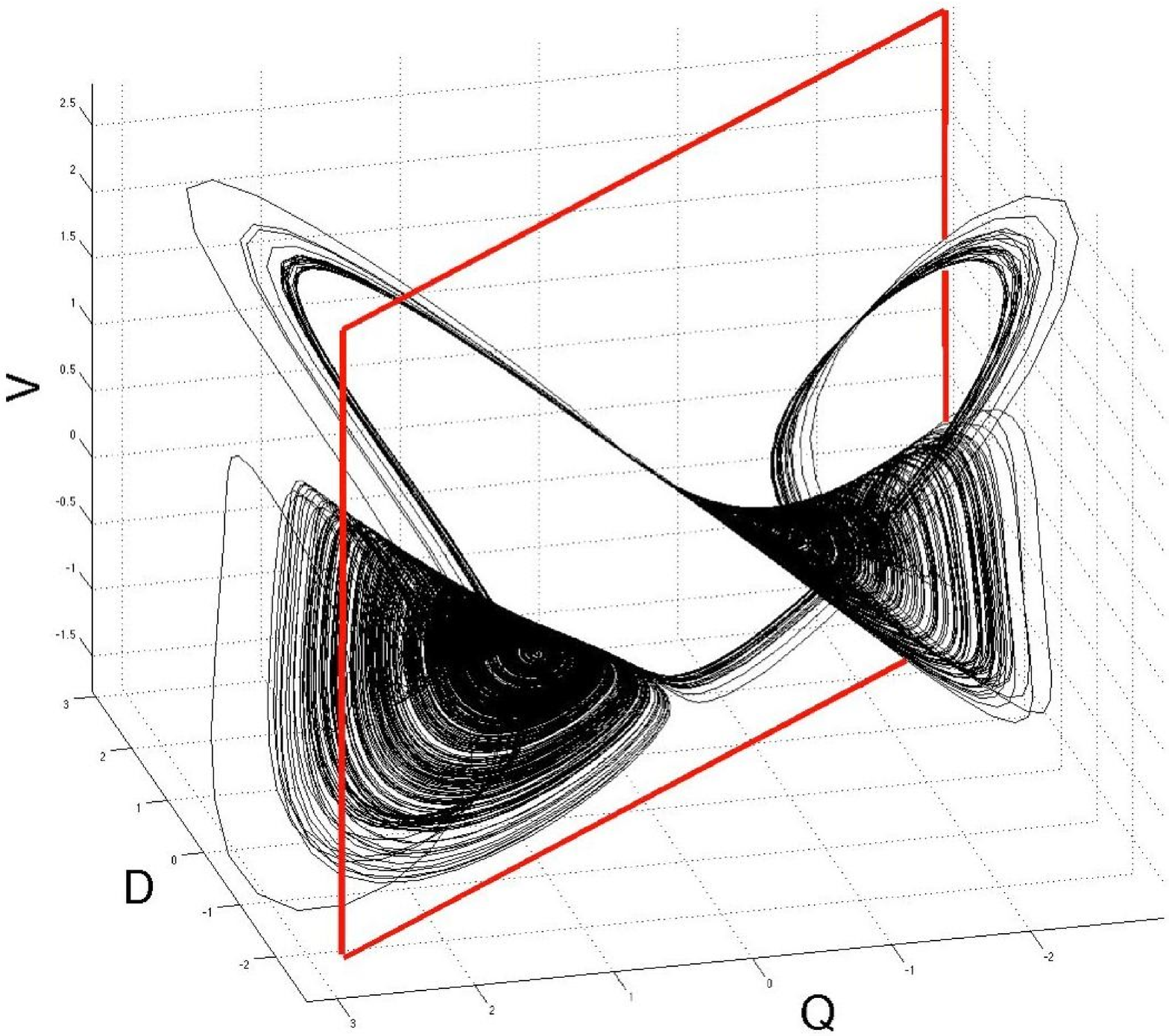}
             \includegraphics[height=62mm]{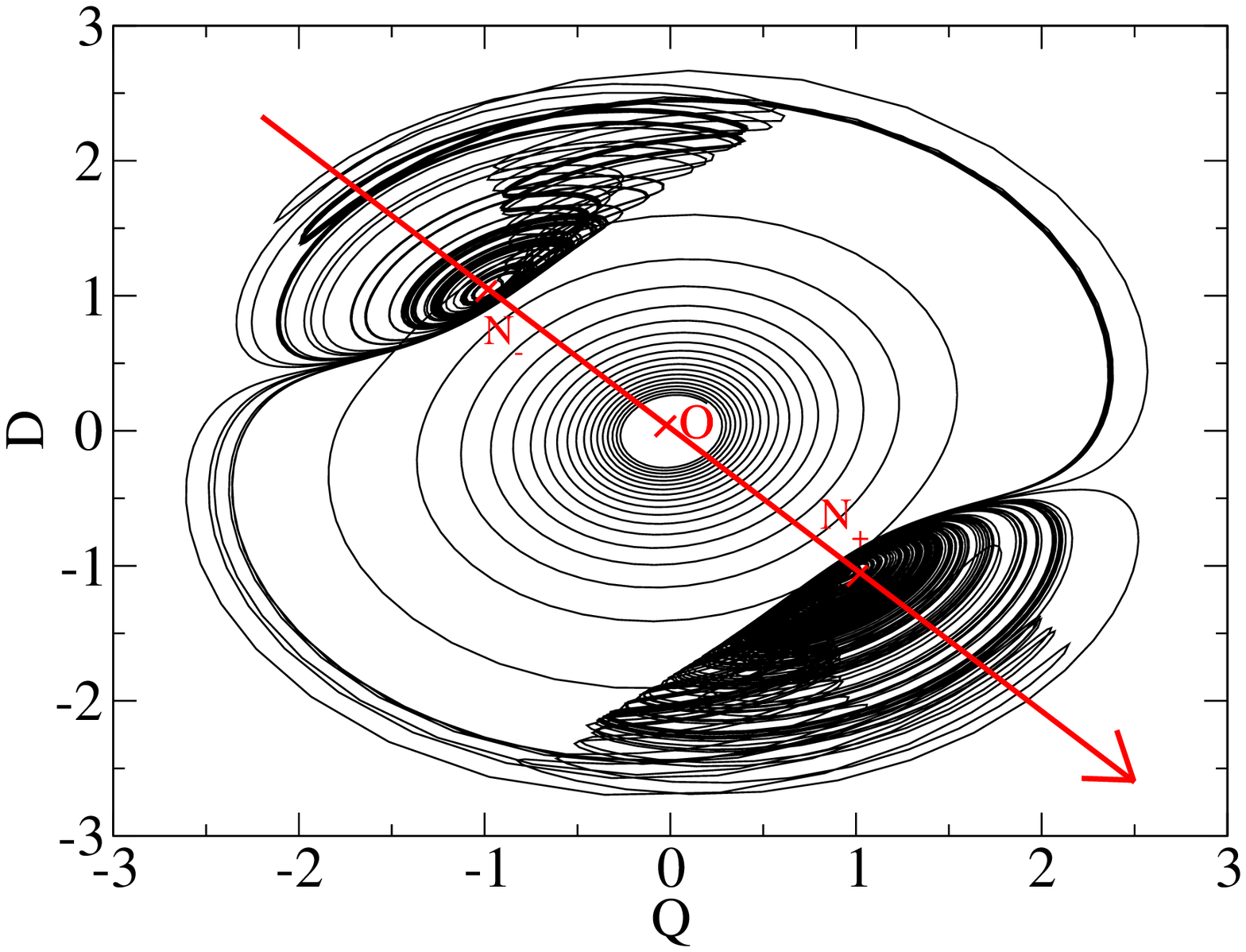}}
		\centerline{
              \includegraphics[height=52mm]{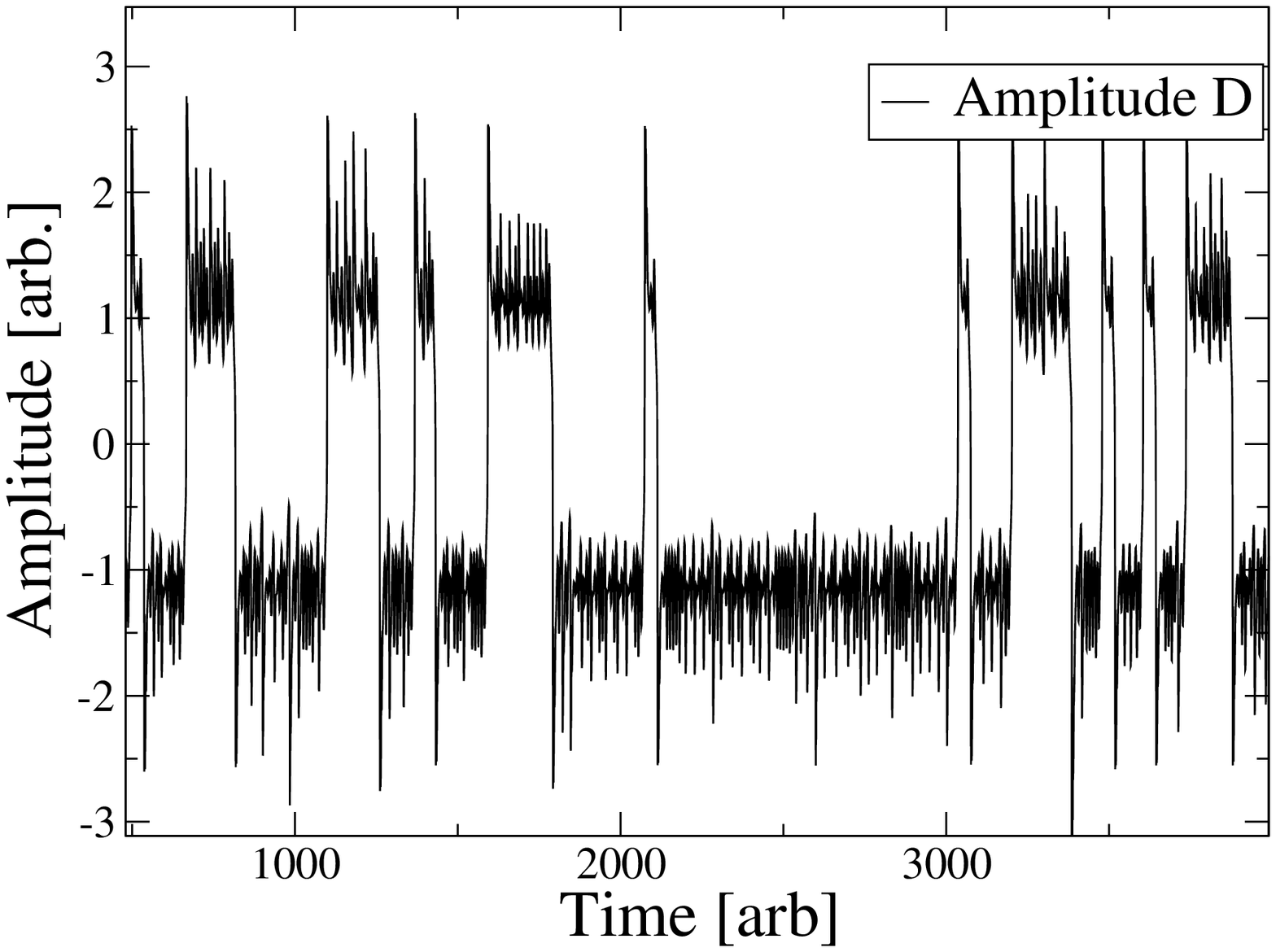}
	      \includegraphics[height=52mm]{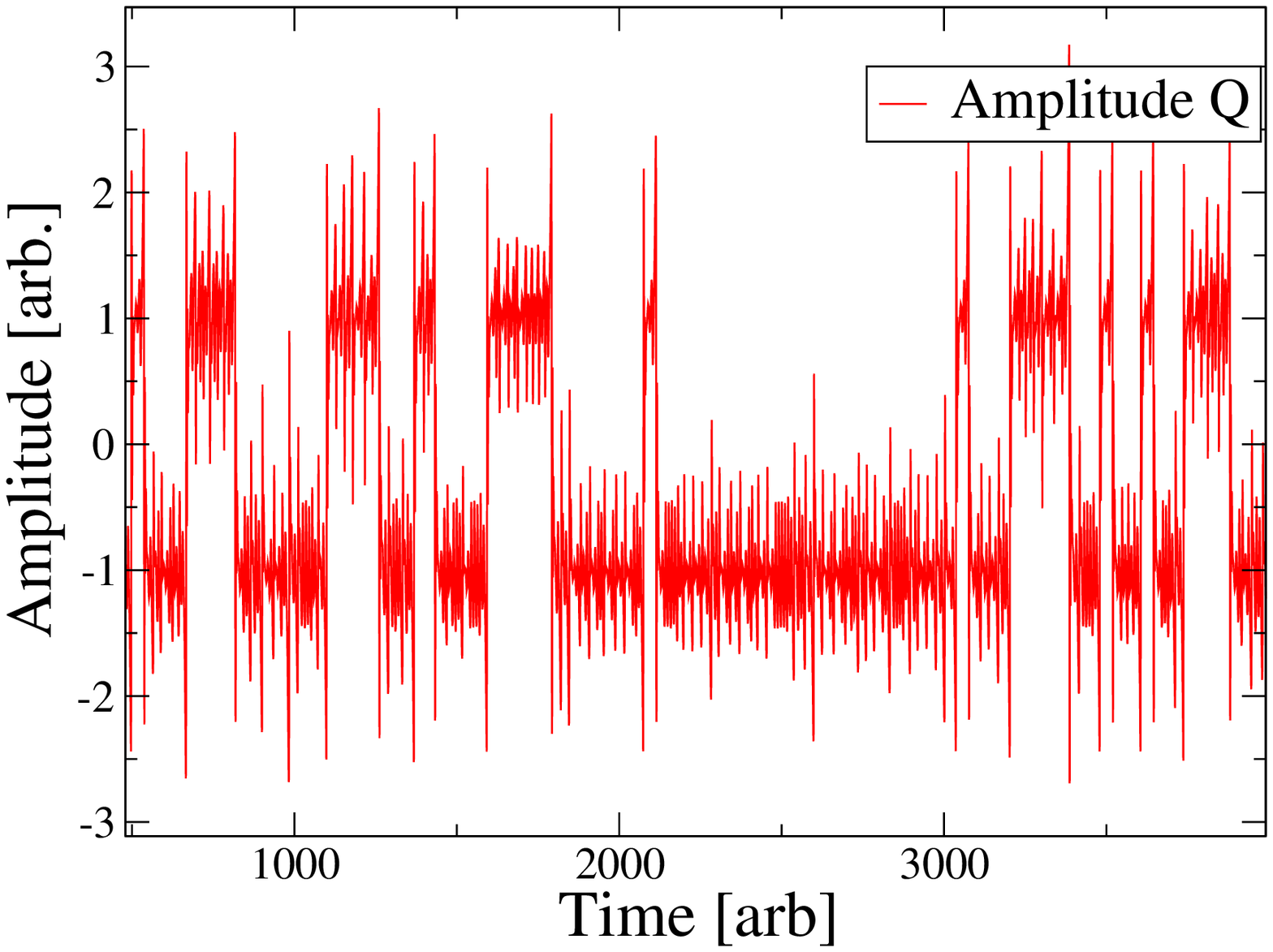}
	      \includegraphics[height=52mm]{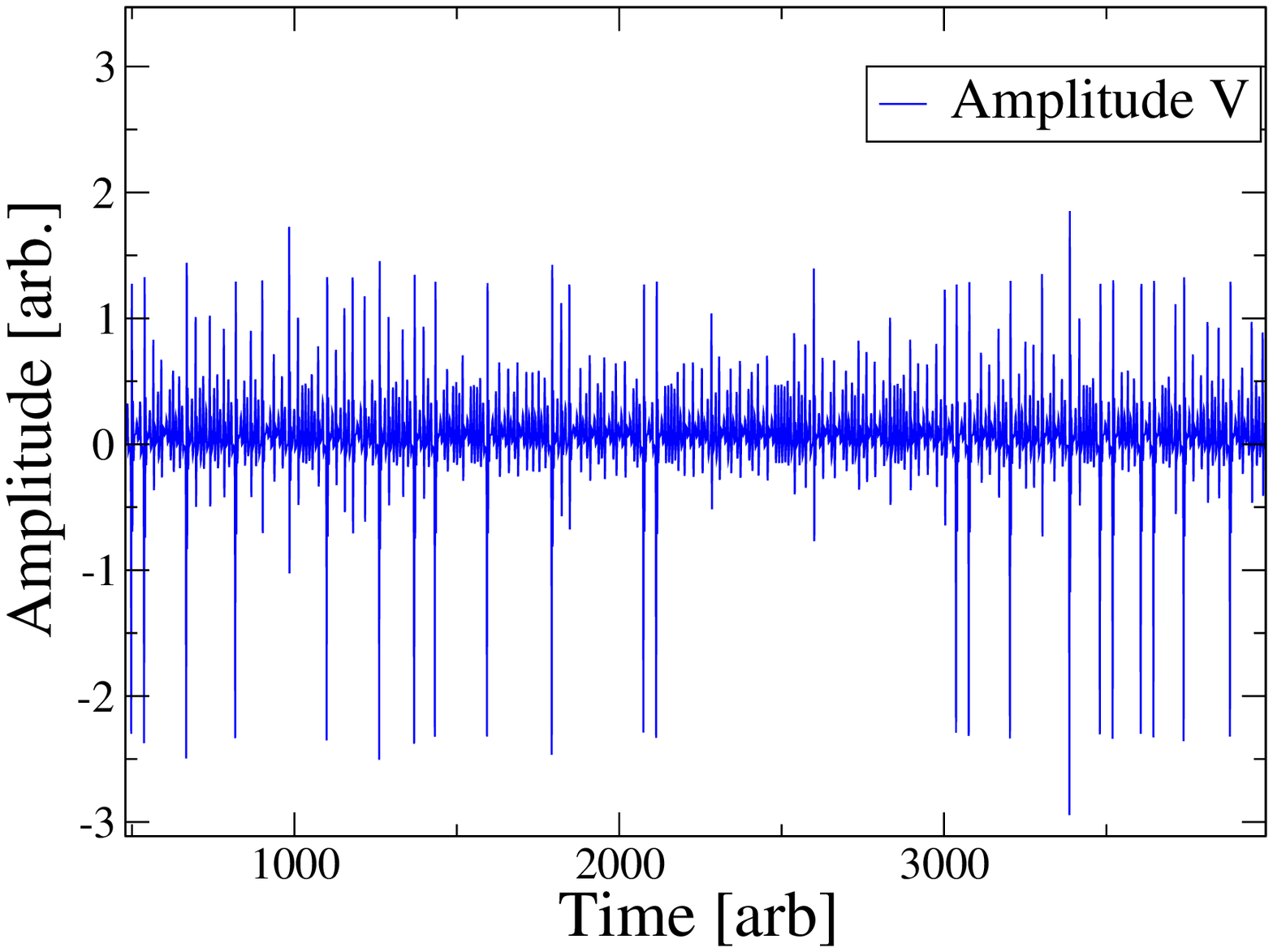}}
\caption{ Numerical integration of the model for $\mu=0.119$,
  $\nu=0.1$ and $\Gamma=0.9$. Poincare section used for iterated map is
  shown in red. Top panel: trajectories in the phase space
  $(D,Q,V)$(left) and in the cut $(D,Q)$ (right). Bottom panel: Time
  evolution of amplitudes $D$(left), $Q$ (middle) and $V$(right).}
\label{mod_attract}
\end{figure*}

In this system of equations, the nonlinear terms are restricted to
quadratic order, and the parameters $\mu$, $\nu$ and $\Gamma$ are
real. Relative signs of the quadratic terms are such that the solution
remains bounded. The divergence of this vector field is equal to
$\mu-\nu-1$, so it is dissipative for $\mu>1+\nu$. This very simple
dynamical system presents similarities with several three-modes models
previously discussed in the literature. In the context of geodynamo
reversals, Rikitake \cite{Rikitake58} proposed a model close to
$(\ref{mod_ampD}-\ref{mod_ampV})$, except that the linear damping of
the mode $V$ was discarded. Moreover, in Rikitake model, $D$ and $Q$
were linearly stable. Nozi\`eres \cite{Noziere78} derived a similar
model, also related to geomagnetic reversals, but obtained from a
truncation of MHD equations, the physical interpretation of this model
being a system of two magnetic components coupled to one velocity
mode. Therefore, in Nozi\`eres model, linear terms represent ohmic
decays and the modes are linearly damped, strongly modifying the
dynamics. Finally, the dynamical system
$(\ref{mod_ampD}-\ref{mod_ampV})$ with $\Gamma=0$ is relevant for
several hydrodynamical problems, and has been extensively studied in a
series of articles by Proctor and Hughes \cite{Hughes90}. Again, we
will see that a non-zero $\Gamma$ is crucial for the dynamics
described in the present paper. Finally, note that the system
possesses the symmetry $(D,Q,V) \rightarrow (-D,-Q,V)$, which is
crucial for the dynamics of reversals reported in section $3$.\\

\section{Stability and fixed points} 
\label{sec:1}

In this section, we first describe the fixed points of the system. A first
trivial fixed point $O$ is related to vanishing amplitudes of $D$ and
$Q$:

\begin{equation}
Q_O=0 \ , \hspace{15pt}
D_O=0 \ , \hspace{15pt}
V_O=\Gamma
\end{equation}
Linearization around this fixed point give the following eigenvalues:

\begin{eqnarray}
\lambda_i&=& \frac{\mu-\nu}{2} \pm
\frac{1}{2}\sqrt{\left(\mu+\nu\right)^2-4\Gamma^2} \ , \hspace{15pt} i=1,2 \\
 \lambda_3&=&-1
\end{eqnarray}
The trivial eigenvalue $\lambda_3$ indicates a stable direction,
independently of the values of $\mu$, $\nu$ and $\Gamma$, and traces
back to the linear damping of the amplitude $V$. When
$\mu+\nu>2\Gamma$, all eigenvalues are real. When $\mu+\nu<2\Gamma$,
eigenvalues $\lambda_1$ and $\lambda_2$ become complex conjugates and
the system undergoes Hopf bifurcation for $\mu>\nu$. Note that this
system exhibits a co-dimension two bifurcation point for
$\mu=\nu=\Gamma$, i.e. the system is close to both a stationary and a
Hopf bifurcation. We will see that the behavior of our dynamical
system strongly depends on the stability of $O$.

Apart from this trivial point, there are four non-trivial fixed points
$M_{\mp}$ and $N_{\mp}$. Two of these solutions only exist when
$\sqrt{\mu \nu}-\Gamma>0$:
\begin{equation}
Q_{M\pm}=\pm\sqrt{\nu-\Gamma\sqrt{\frac{\nu}{\mu}}}
\end{equation}
\begin{equation}
D_{M\pm}=\pm\sqrt{\mu-\Gamma\sqrt{\frac{\mu}{\nu}}}
\end{equation}
\begin{equation}
V_{M\pm}=\sqrt{\mu\nu}
\end{equation}
\\

The other fixed points are :
\begin{equation}
Q_{N\pm}=\pm\sqrt{\nu+\Gamma\sqrt{\frac{\nu}{\mu}}}
\end{equation}
\begin{equation}
D_{N\pm}=-\pm\sqrt{\mu+\Gamma\sqrt{\frac{\mu}{\nu}}}
\end{equation}
\begin{equation}
V_{N\pm}=-\sqrt{\mu\nu}
\end{equation}

Linearization of the system around the fixed points $M_{\pm}$ gives
the following characteristic relation:\\

\begin{equation}
\lambda^3+\lambda^2\left(1+\nu-\mu\right) + \lambda\left(\Gamma\frac{(\nu-\mu)}{\sqrt{\mu\nu}} \right)+ 4\mu\nu-4\Gamma\sqrt{\mu\nu}=0 \, .
\end{equation}

Similarly, for $N_{\pm}$ we have:\\
\begin{equation}
\lambda^3+\lambda^2\left(1+\nu-\mu\right) - \lambda\left(\Gamma\frac{(\nu-\mu)}{\sqrt{\mu\nu}} \right)+ 4\mu\nu+4\Gamma\sqrt{\mu\nu}=0 \, .
\end{equation}

In the following, we describe some of the chaotic behaviors generated
by the system of equations $(\ref{mod_ampD}-\ref{mod_ampV})$, by
exploring the parameter space while $\Gamma$ is kept fixed (in
practice $\Gamma=0.9$). We consider only situations where the
linearization around non-trivial fixed points gives one real
eigenvalue, the two others being complex conjugates. The real
eigenvalue is always negative, ensuring that the fixed points have a
stable direction. The real part of the complex conjugate eigenvalues
will depend on the value of $\mu$ and $\nu$.

\begin{figure}[ht!]
	\centerline{\includegraphics[width=0.85\columnwidth]{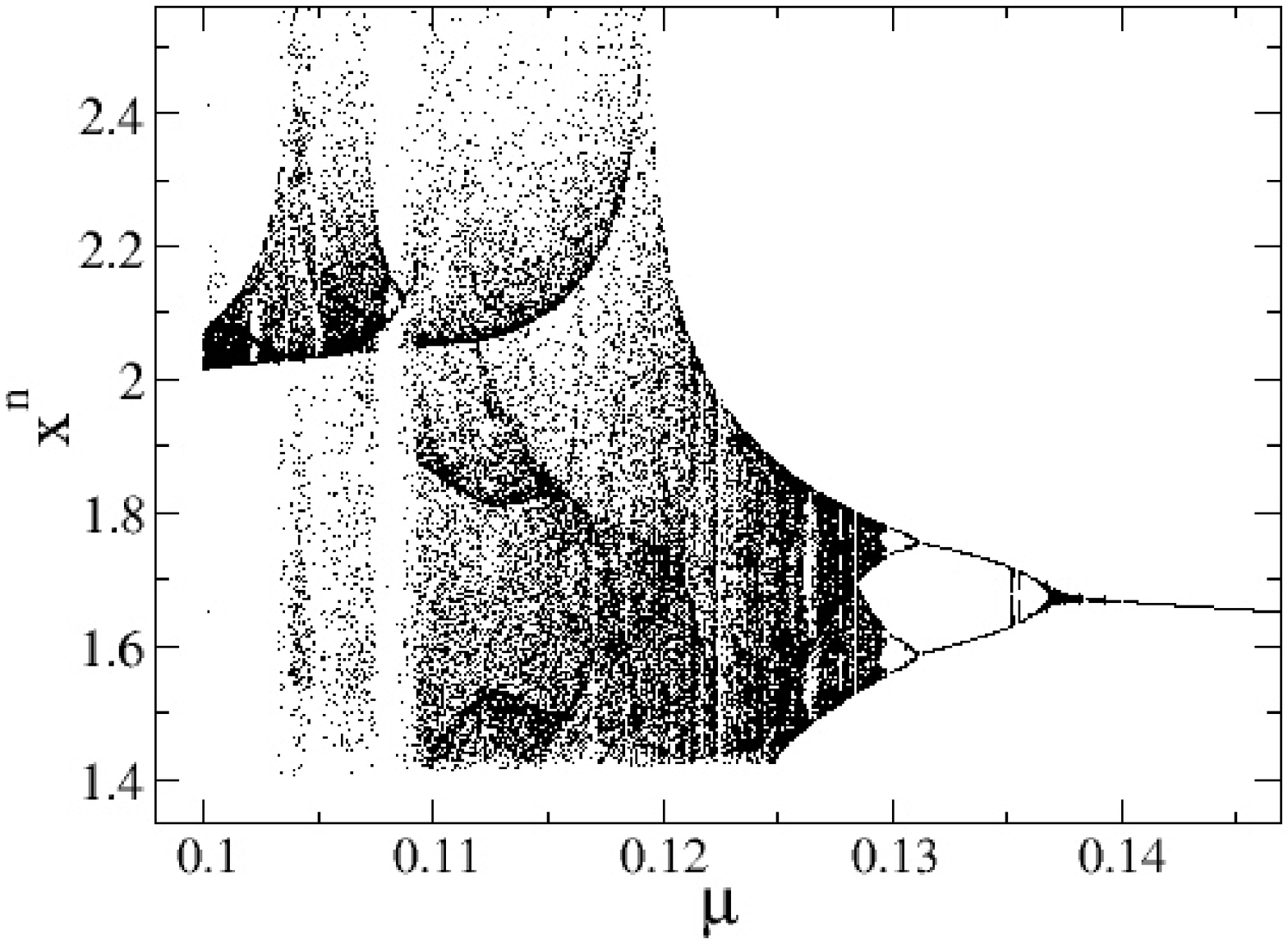}}
\vspace*{-3mm}
        \centerline{\includegraphics[width=0.85\columnwidth]{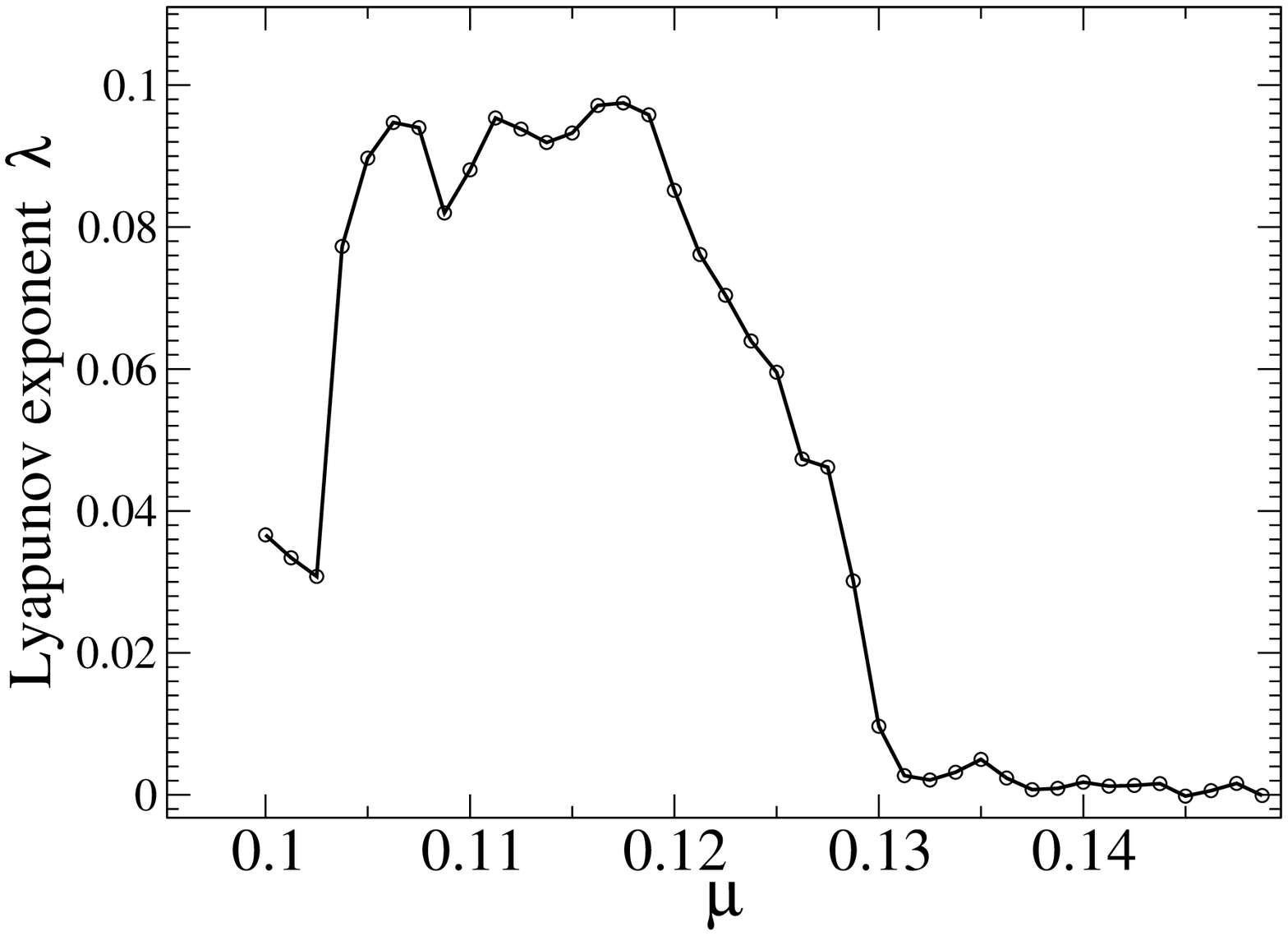}}
\caption{ Top: Bifurcation diagram of the return map $x_n$ as a
  function of $\mu$. When $\mu$ is decreased from $0.15$ to $0.1$, the
  transition to chaos is achieved through a period doubling
  cascade. Bottom: Evolution of the Lyapunov exponent $\lambda$ of the
  system when $\mu$ is decreased. The successive period doubling
  events lead to a chaotic state, characterized by an increase of
  $\lambda$.}
\label{mod_chaosbif}
\end{figure}

Among the wide variety of dynamical regimes observed in this simple
system, a particularly interesting chaotic behavior can be observed
for some parameters (see for instance the chaotic attractor shown in
figure \ref{mod_attract}, obtained for $\mu=0.119$, $\nu=0.1$ and
$\Gamma=0.9$). For these parameter values, the system explores the
phase space around one of the fixed points $N$ for a given time, and
chaotically switches from one solution to its opposite, leading to
chaotic reversals of amplitudes $D$ and $Q$ ($V$ does not change
sign). Although the evolution during a given polarity is strongly
chaotic, reversals exhibit very robust characteristics.  In order to
understand the mechanisms leading to such dynamics, in particular how
the system becomes chaotic, it is useful to keep parameters $\Gamma$
and $\nu$ to these values ($\Gamma=0.9$ and $\nu=0.1$), while only
$\mu$ is varied. In addition, the dimension of the problem can be
reduced by means of a Poincare map in the phase space
\cite{GuckHolmes}. As shown in figure \ref{mod_attract}, Poincare map
is defined as the surface orthogonal to the $(D,Q)$ plane which
contains the line joining the two fixed points $N_+$ and $N_-$ (this
line provides a graduated axis with $O$ the origin).  Our first return
map is then defined by the intersection $x_{n+1}$ between trajectories
of the system (projected in the $(D,Q)$ plane) and this line, as a
function of the previous intersection $x_n$. Figure \ref{mod_chaosbif}
shows a bifurcation diagram of $x_n$ as $\mu$ is decreased from $0.15$
to $0.1$.

When $\mu>0.2$, two stationary solutions are obtained, corresponding
to the fixed points $N_+$ and $N_-$. For $\mu<0.2$, these fixed points
loose their stability and a stable periodic orbit is created around
each fixed points, corresponding to stable solutions in the iterated
map.
Below $\mu=0.138$, a series of pitchfork bifurcations is observed in
the return map of figure \ref{mod_chaosbif}-top, due to the
successive period doubling bifurcations of the periodic orbit
surrounding $N_+$. Because of the symmetry of the system, a similar
transition occurs around $N_-$.

\begin{figure}[ht!]	
		\centerline{\includegraphics[width=0.85\columnwidth]{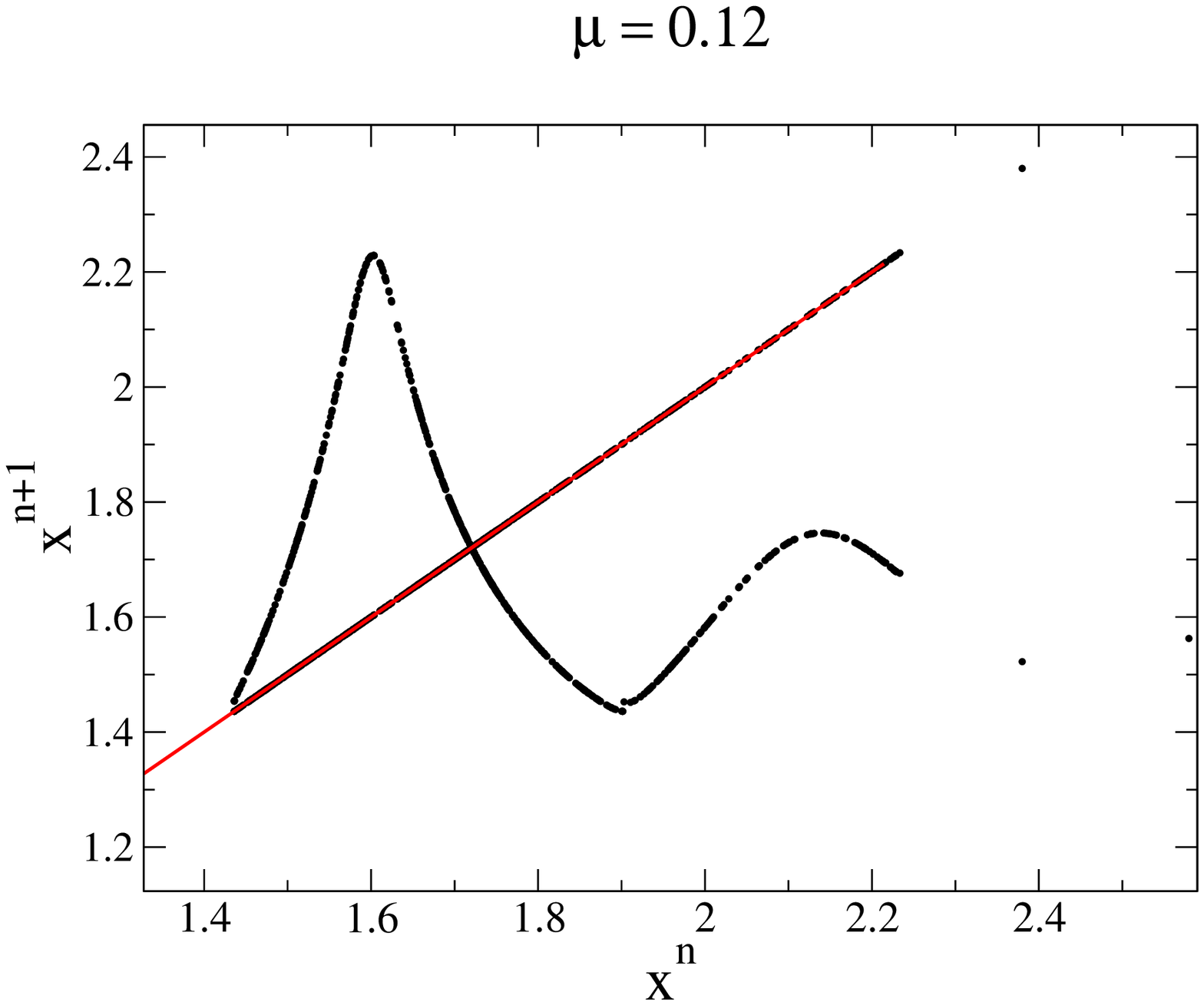}}
\vspace*{-3mm}
		\centerline{\includegraphics[width=0.85\columnwidth]{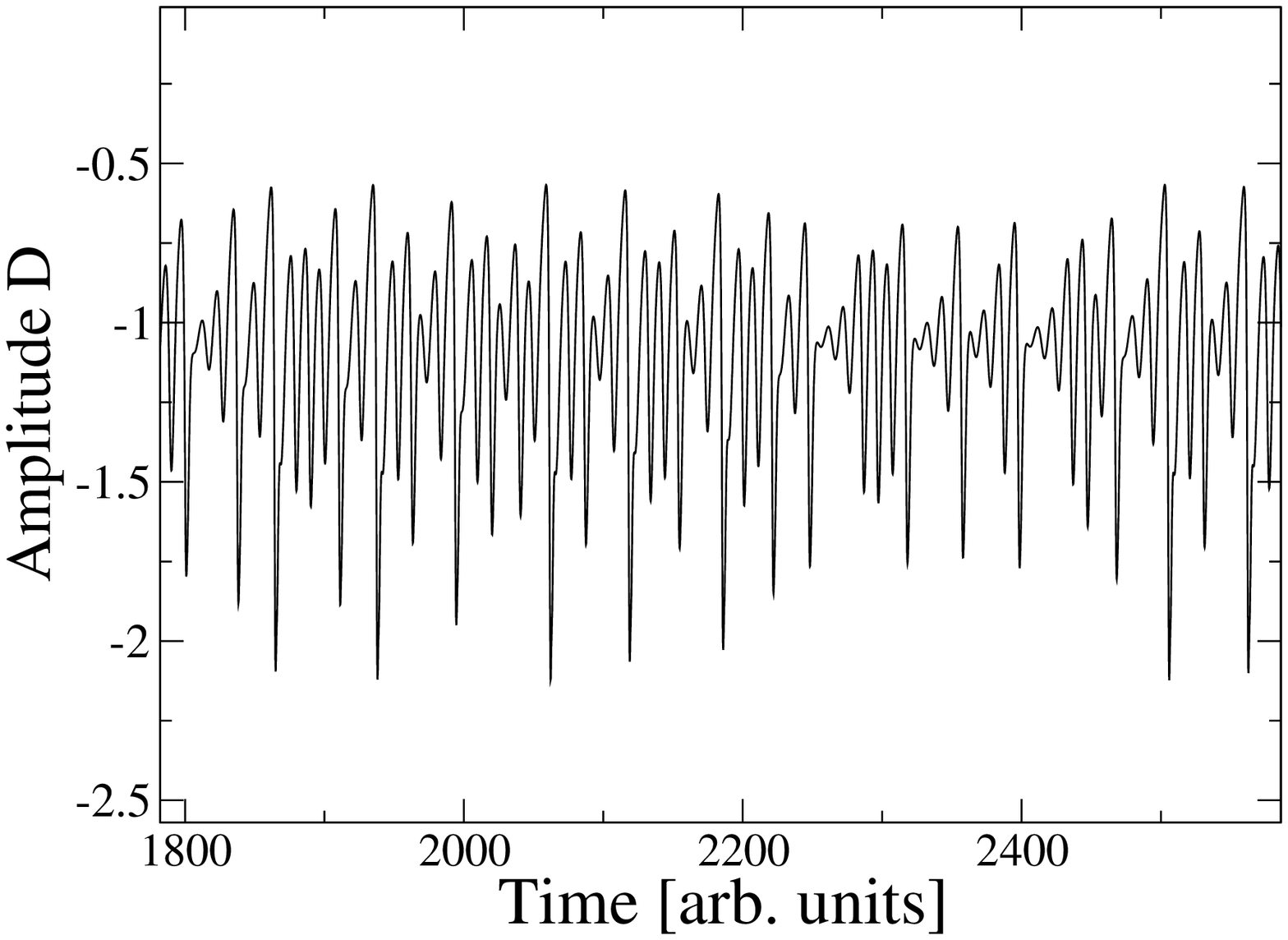}}
\caption{ Numerical integration of the model for $\mu=0.12$, $\nu=0.1$
  and $\Gamma=0.9$. These parameters correspond to a fully chaotic
  state, with trajectories fluctuating around one of the fixed
  points. Top: discrete map. Bottom: time evolution of $D$. Initial
  conditions are chosen in the basin of attraction of $N_+$.}
\label{mod_mu0p12}
\end{figure}

We thus see that our simple dynamical system exhibits a classical
route to chaos, by period doubling bifurcations. This scenario is
observed in several discrete applications, like the Henon map or the
logistic map. Note that the transition to chaos is very sharp, with a
Lyapunov exponent $\lambda$ (characterizing the rate of separation of
infinitesimally close trajectories) which rapidly tends to $0.1$ at
$\mu\approx 0.13$ (see figure \ref{mod_chaosbif}-bottom).

Figure \ref{mod_mu0p12} shows trajectories for $\mu=0.12$, for which a
chaotic attractor is obtained. On the iterated map, the system
randomly explores the interval $1.4<x_n<2.3$ (figure \ref{mod_mu0p12},
top), defining a typical size of the chaotic attractor, bounded in the
phase space $(D,Q,V)$. It is important to note that symmetries of our
dynamical system impose that a similar attractor is obtained in the
vicinity of the opposite fixed point $N_-$. However, for $\mu=0.12$,
these two symmetric attractors are disconnected, and there are no
reversals (figure \ref{mod_mu0p12}, bottom). This is a noticeable
difference with other deterministic models for reversals. For
instance, in the Lorenz model, a period doubling route to chaos can
also be observed \cite{Smith98}, but in that case, the initial periodic
orbit is a cycle surrounding the fixed points of the system, and
therefore already involving reversals of the amplitudes.  In the next
section, we describe how a further decrease of $\mu$ yields chaotic
reversals of $D$ and $Q$, very different from reversals obtained in
Lorenz or Rikitake models.

\section{Crisis-induced intermittency and reversals}
\label{sec:2}
\begin{figure}[ht!]
		\centerline{\includegraphics[width=0.85\columnwidth]{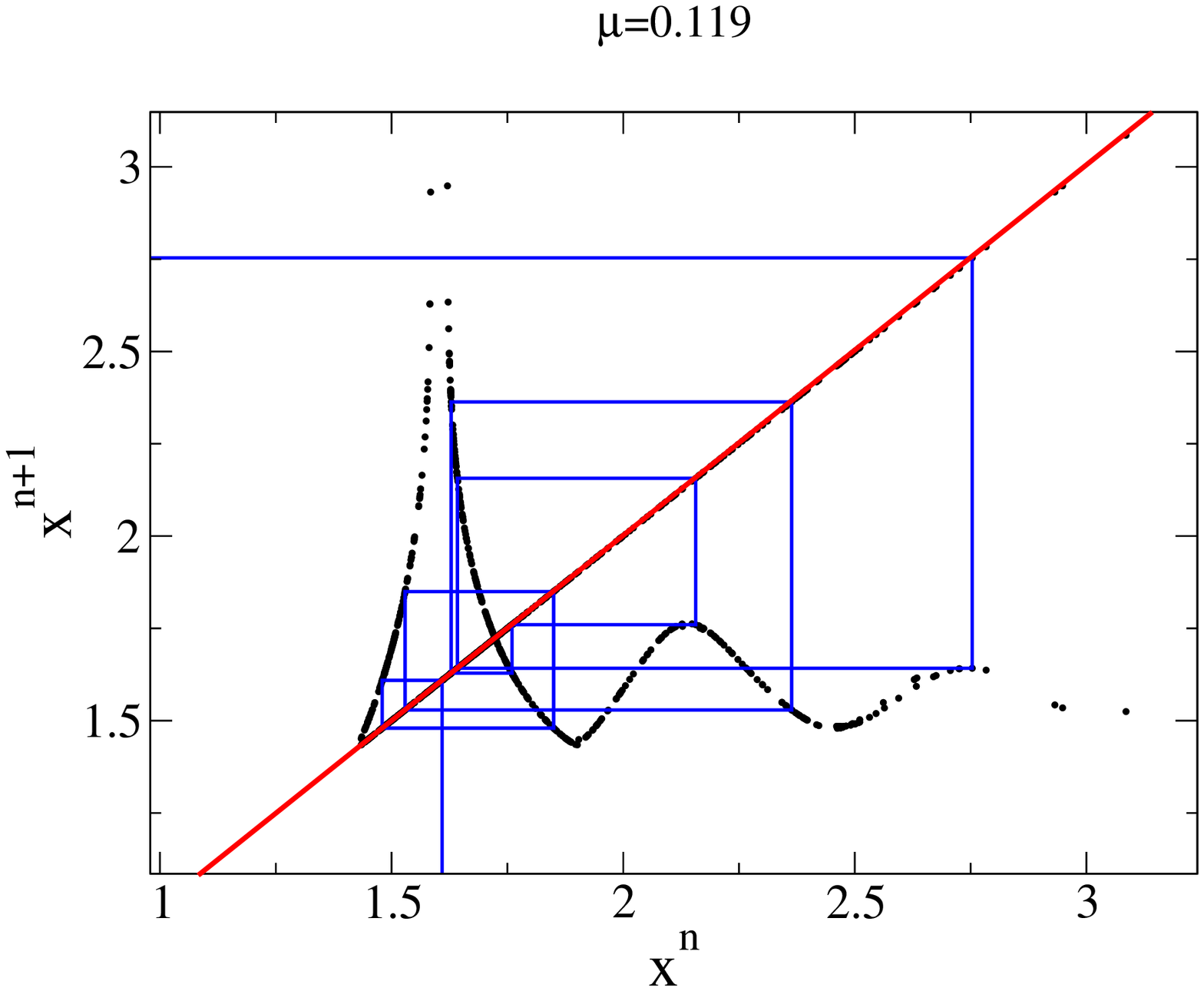}}
\vspace*{-3mm}
		\centerline{\includegraphics[width=0.85\columnwidth]{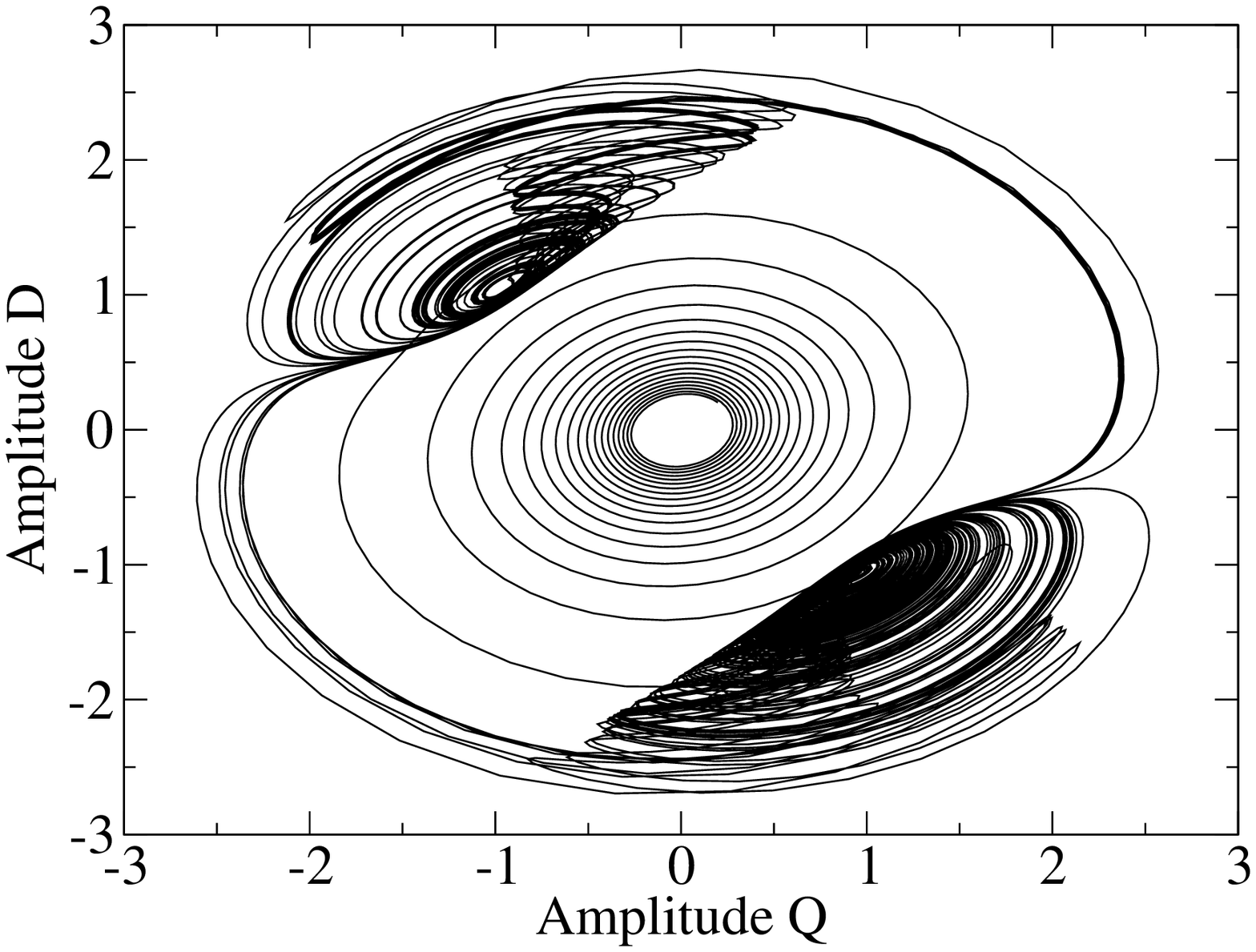}}
\caption{ Dynamics of the model for $\mu=0.119$, $\nu=0.1$ and
  $\Gamma=0.9$. Discrete application (top) is very similar to the one
  obtained for $\mu=0.12$, except for a small region connected to the
  opposite fixed point. Bottom: evolution of the system in the phase
  space $(D,Q)$.}
\label{mod_mu0p119}
\end{figure}
For $\mu=0.12$, the system chaotically fluctuates around a fixed
point, but without reversing. From this state, a slight decrease of
$\mu$ yields a transition to a dynamics of reversals between two
symmetric states with well defined values. Figure \ref{mod_mu0p119}
shows the evolution of the model for $\mu=0.119$, in such a reversing
regime.  Note that the behavior close to one of the fixed points is
similar to the one generated for $\mu=0.2$, i.e without
reversals. However, for $\mu=0.119$, the two attractors are now
connected, allowing chaotic reversals between the two opposite states.

Changing one of the parameter of the system (here $\mu$) yields a
discontinuous change of the structure of the attractor. This typical
behavior of dissipative systems is known as a crisis phenomenon. The
crisis behavior of dynamical systems has been discovered and
extensively studied by Grebogi, Ott and York \cite{Grebogi82},
\cite{Grebogi87}, mainly in the context of discrete maps. Following
these authors, a crisis is defined as ``\emph{a collision between a
  chaotic attractor and a coexisting unstable fixed point or periodic
  orbit}''.

Typically, in a crisis scenario, modifications of the structure of the
attractor are of three kinds: in a first case, the chaotic attractor
is suddenly destroyed at the critical point. In a second case, the
size of the attractor undergoes an abrupt increase. In a last case, two
attractors or more collide and a larger attractor is created. In each
of these three cases, the structural change is related to a collision
with an unstable attractor (chaotic or not), and it is possible to
characterize the phenomena with a typical time $\tau$. This $\tau$
measures the finite time during which the system behaves as before the
crisis, and then rapidly leaves this region of chaotic transient.

In our dynamical system, the apparition of chaotic reversals seems to
belong to the third class, i.e. the merging of two distinct
attractors. Let us define $\mu_c$ as the critical value of $\mu$ for
which reversals appear. When $\mu>\mu_c$, the two attractors close to
each of the fixed points possess their own basin of attraction,
separated by a boundary basin. Depending on the initial conditions,
trajectories tend to one attractor or the other, and are trapped in
the basin. Because of the symmetry of the system, when $\mu=\mu_c$,
the two attractors simultaneously collide the boundary basin
separating them, which allows a connection between both attractors,
through some unknown unstable periodic orbits of the attractors. As a
consequence, for $\mu=\mu_c$, the system spends a long time exploring
chaotically the vicinity of one of the fixed point before it suddenly
leaves this region and fluctuates around the opposite solution. The
two attractors being symmetrical, the process then repeats and chaotic
reversals are generated. We can thus characterize the dynamics by a
time $\tau$ which is naturally defined as the average duration between
two successive reversals, hereafter referred as Waiting Time Duration
(WTD).

In general, Grebogi et al have shown that $\tau$ follows a power law
\begin{equation}
\tau\sim \left(\mu-\mu_c\right)^{-\gamma}
\label{mod_taulaw}
\end{equation}

Using numerical simulations, we determined numerically this law. For
each value of $\mu$, $\tau$ is obtained by averaging all the WTDs
obtained in a time series containing one thousand reversals.

\begin{figure}[ht!]
		\centerline{\includegraphics[width=0.85\columnwidth]{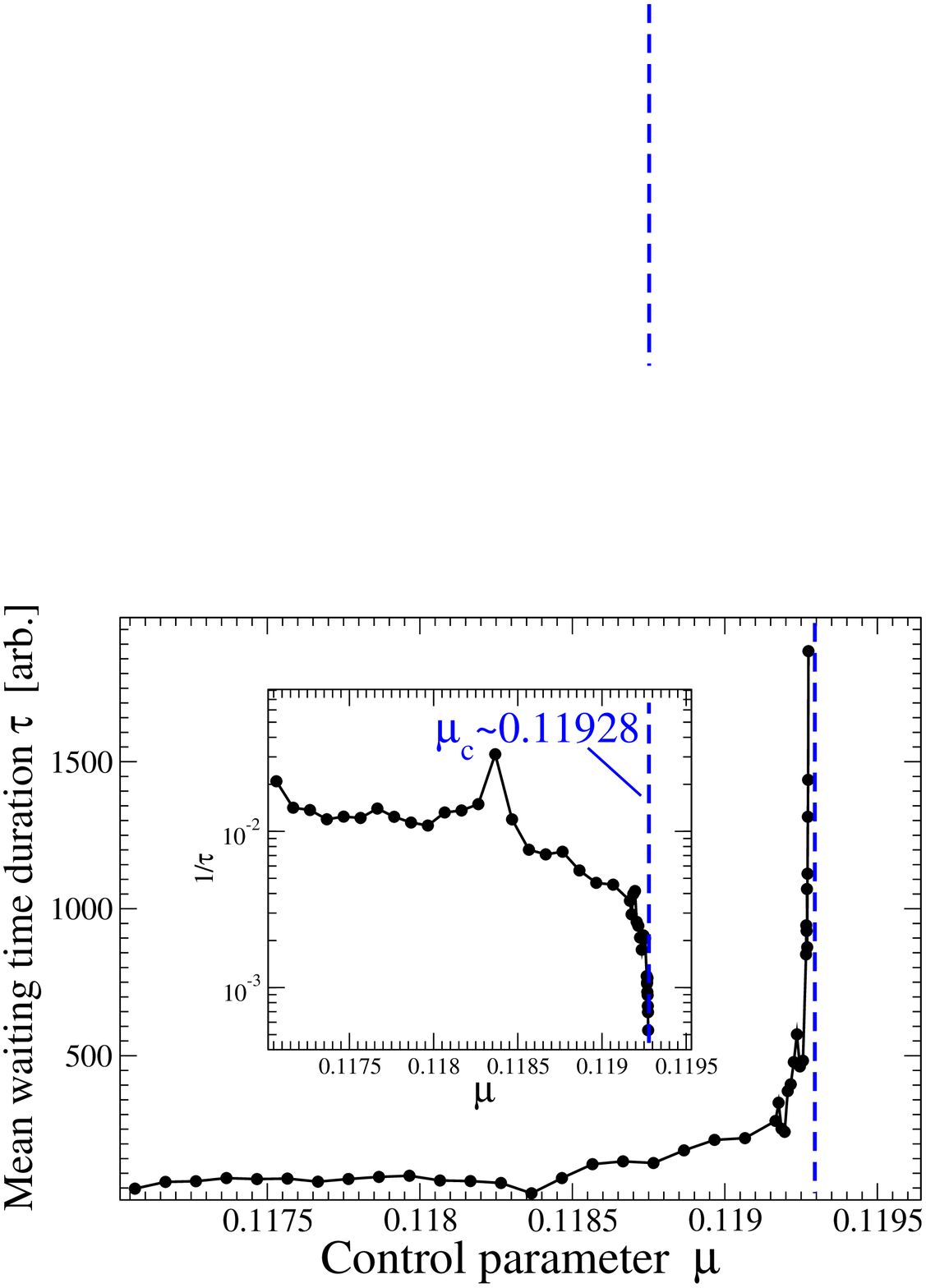}}
\vspace*{-3mm}
                \centerline{\includegraphics[width=0.85\columnwidth]{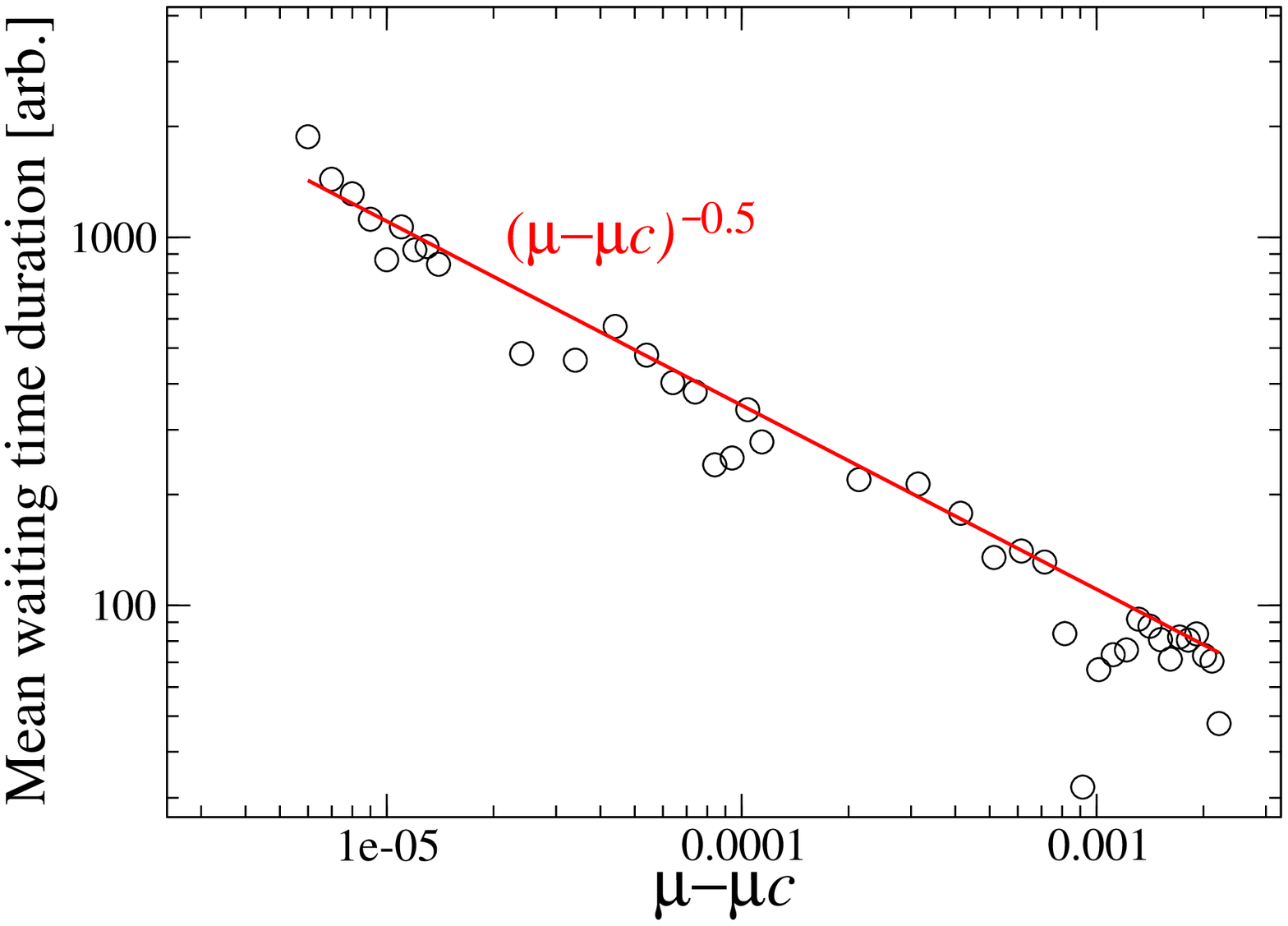}}
\caption{ Evolution of the characteristic waiting time duration
  between two successive reversals when the critical point of the
  crisis is approached. Top: $\tau$ as a function of $\mu$,
  suggesting $\mu_c=0.11928$. Inset shows $1/\tau$($\mu$). Bottom:
  Evolution in logarithmic scale as a function of
  $\mu-\mu_c$. Numerical simulations suggest an exponent close to
  $\gamma \sim 1/2$.}
\label{mod_tau0p119}
\end{figure}

Figure \ref{mod_tau0p119}-top shows the averaged waiting time
duration $\tau$ between two successive reversals as a function of
$\mu$. The curve is diverging for $\mu_c \sim 0.11928$. In practice,
this critical value has been obtained by noting that no reversals
occurred after an integration time $t>1.10^5$, whereas reversals were
observed for $\mu=0.119275$.

Figure \ref{mod_tau0p119}-bottom shows how the waiting time durations
evolve as the critical point is approached. There is little doubt that
this curve follows a power law. However, even if our numerical
simulations suggest a critical exponent $\gamma \simeq 1/2$, it is
difficult to conclude on a precise exponent, since a small error on
the determination of $\mu_c$ significantly modifies the measured power
law.

In some idealized cases, Grebogi et al. \cite{Grebogi87} have shown
that it is possible to obtain the critical exponent $\gamma$
theoretically and to express it as a function of the eigenvalues of
the periodic orbits involved in the crisis.  In the case of discrete
maps:
 \begin{equation}
\gamma={1\over 2}+(\ln|\alpha_1|)/\ln|\alpha_2|\
\end{equation}
where $\alpha_1$ et $\alpha_2$ are respectively eigenvalues of the
unstable and stable manifolds of the two periodic orbits considered.
In principle, it should be possible to use this formula to obtain
$\gamma$ theoretically from the Poincare map. Unfortunately, the
method involves calculating numerically the stable and unstable
manifolds of the orbits mediating the crisis, leading to an error bar
on the estimate of $\gamma$ too large to be compared to the exponent
computed from the waiting time durations. Although the value of
$\gamma$ seems very close to the theoretical lower bound, our
numerical results are not accurate enough to conclude.

\begin{figure*}[ht!]
	\centerline{\includegraphics[width=0.85\columnwidth]{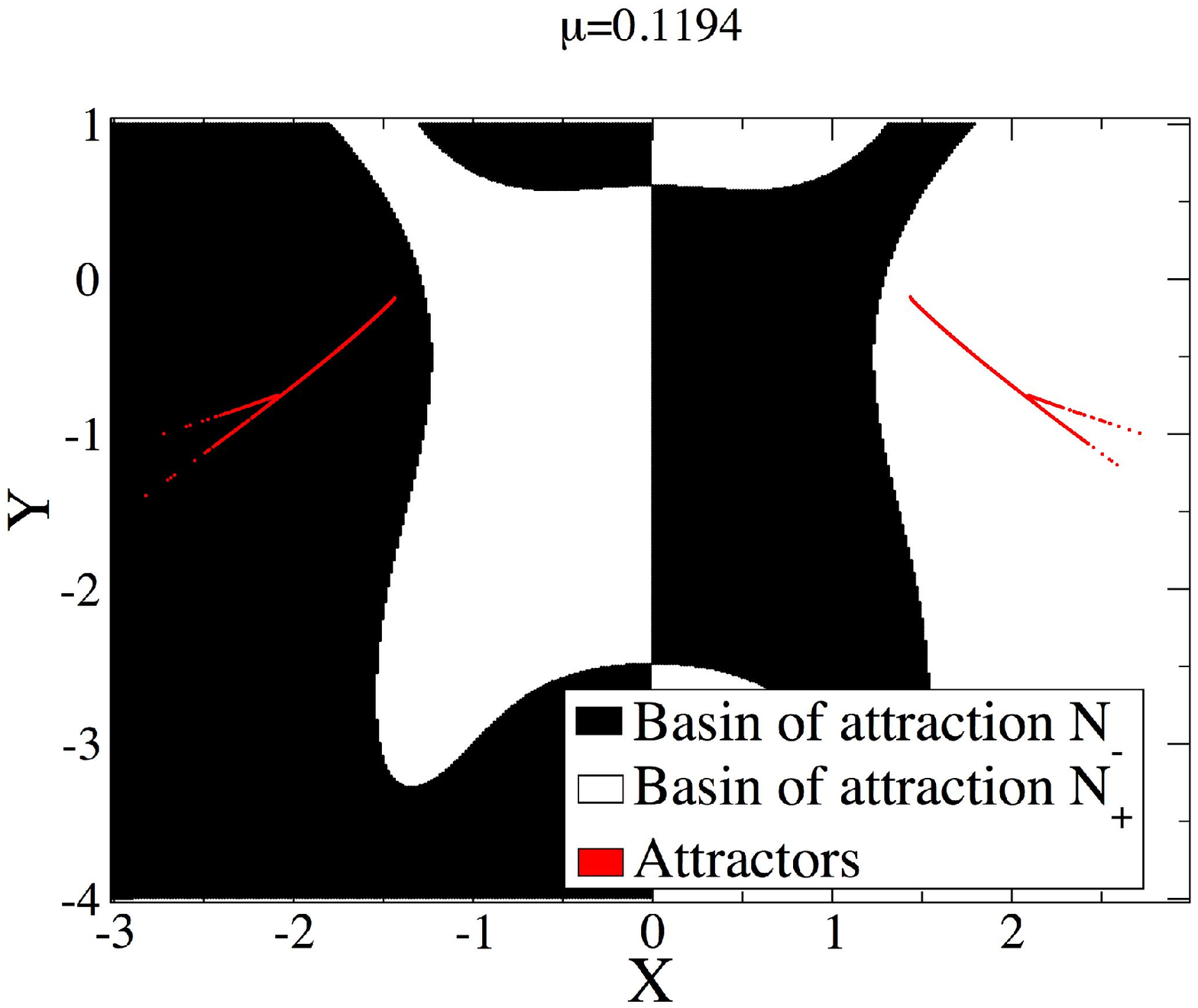}
                    \includegraphics[width=0.85\columnwidth]{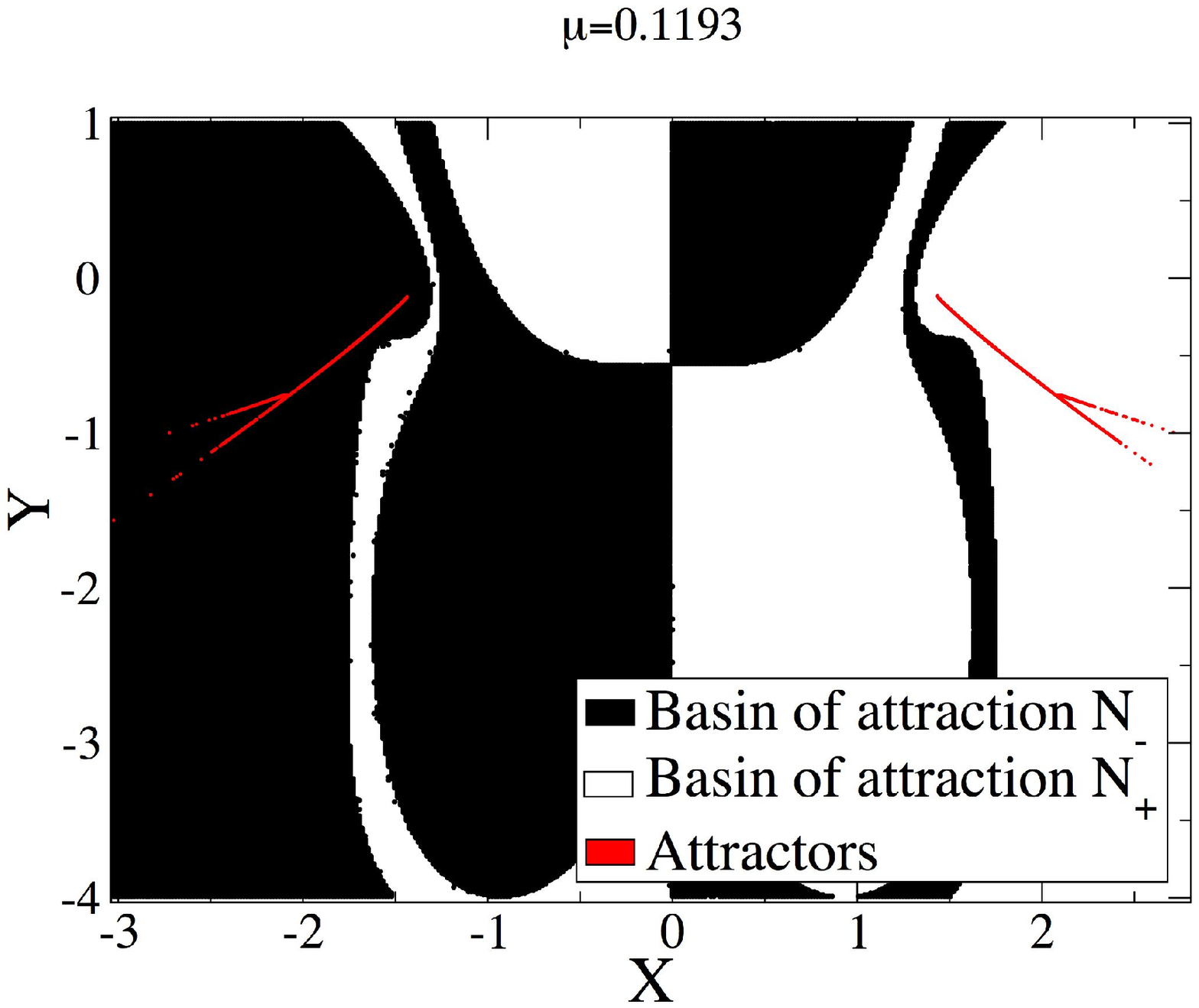}}
\vspace*{-3mm}
        \centerline{\includegraphics[width=0.85\columnwidth]{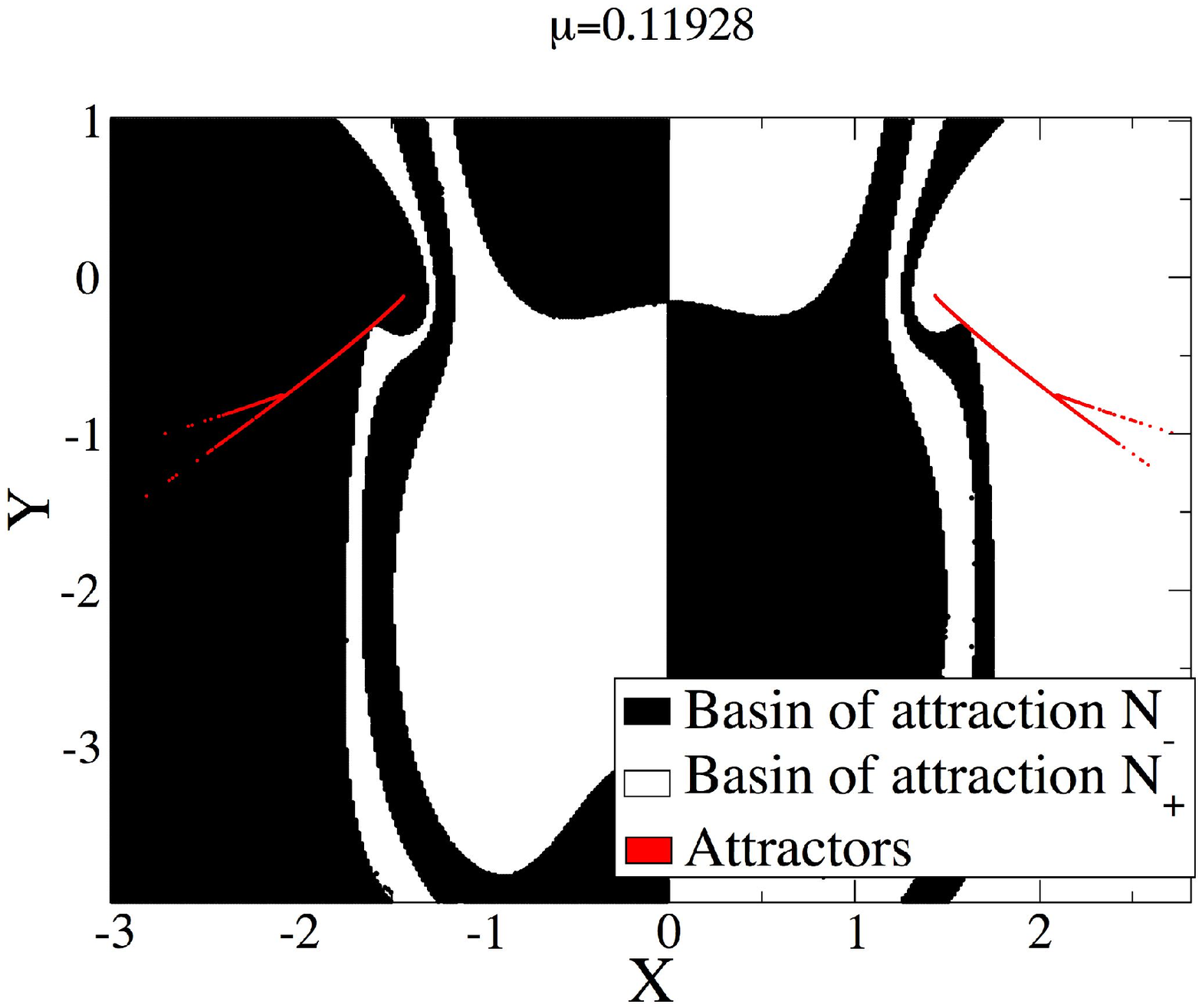}}
\caption{Basin of attraction of chaotic attractors projected in the
  Poincare section. White and black colors distinguish the basins of
  attraction of each attractors. Attractors are shown in red. Top:
  before the crisis, for $\mu=0.1194$ (left) and $\mu=0.1193$
  (right). Bottom: For $\mu=0.11928\approx \mu_c$, we observe a
  collision between attractors at the basin boundaries, illustrating
  the boundary crisis. }
\label{mod_bassin}
\end{figure*}

This relation between the exponent and the eigenvalues of the periodic
orbits relies on the assumption that the crisis is of the
\emph{heteroclinic tangent} type, which means that it is induced when
the unstable manifold of a periodic orbit $(A)$ becomes tangent to the
stable manifold of a second unstable periodic orbit $(B)$ (located
outside the basin of attraction before the crisis). One particularity
of our dynamical system is that the orbit $(B)$ destroying the first
attractor belongs to the basin of attraction of the second
attractor. Because of the symmetries, identical connections are
achieved with both attractors, allowing the system to chaotically come
back to the attractor it previously leaved.

From the Poincare section defined in figure \ref{mod_attract}, it is
possible to visualize the collision leading to the crisis, and figure
\ref{mod_bassin} shows the basin of attraction of the system in the
Poincare section $(X,Y)$.
Theses basins are numerically obtained by using a grid of $(500\times
500)$ initial conditions $(X_0,Y_0)$: the point $(X_0,Y_0)$ is colored
in black (respectively in white) if the computed trajectory starting
from this initial condition ultimately tends to the attractor
surrounding $N_-$ (respectively $N_+$). Figure \ref{mod_bassin}(a)
shows the system before the crisis, for $\mu=0.1194>\mu_c$. In red are
points corresponding to the intersection of trajectories of the
attractor with the Poincare section. Note that the two attractors are
distinct. Black and white regions clearly indicate symmetric basins of
attraction, and are such that the two attractors are connected only to
their own basin. Figure \ref{mod_bassin}(b) corresponds to a slightly
smaller value, $\mu=0.1193>\mu_c$, showing that a change of $\mu$
toward $\mu_c$ yields a change of shape of the basins of attraction
and an increase of their size. Finally, for $\mu=0.11928\approx
\mu_c$, each of the chaotic attractor collides with the boundary of
the basin of attraction of the symmetric attractor. A look at the
trajectories in the phase space indicates that the orbit involved in
this boundary crisis seems to be an unstable periodic orbit encircling
the trivial point $O$. Note indeed that this kind of reversals are
only observed when $O$ is an unstable focus.  It is interesting to
note that when these reversals appear, the largest Lyapunov exponent
is not modified, since reversals are due to the fusion of two
identical chaotic attractors nearly unchanged by the crisis event. In
this regime, we thus obtain a Lyapunov exponent $\lambda \simeq
0.1$.\\

\begin{figure}[ht!]
		\centerline{\includegraphics[width=0.85\columnwidth]{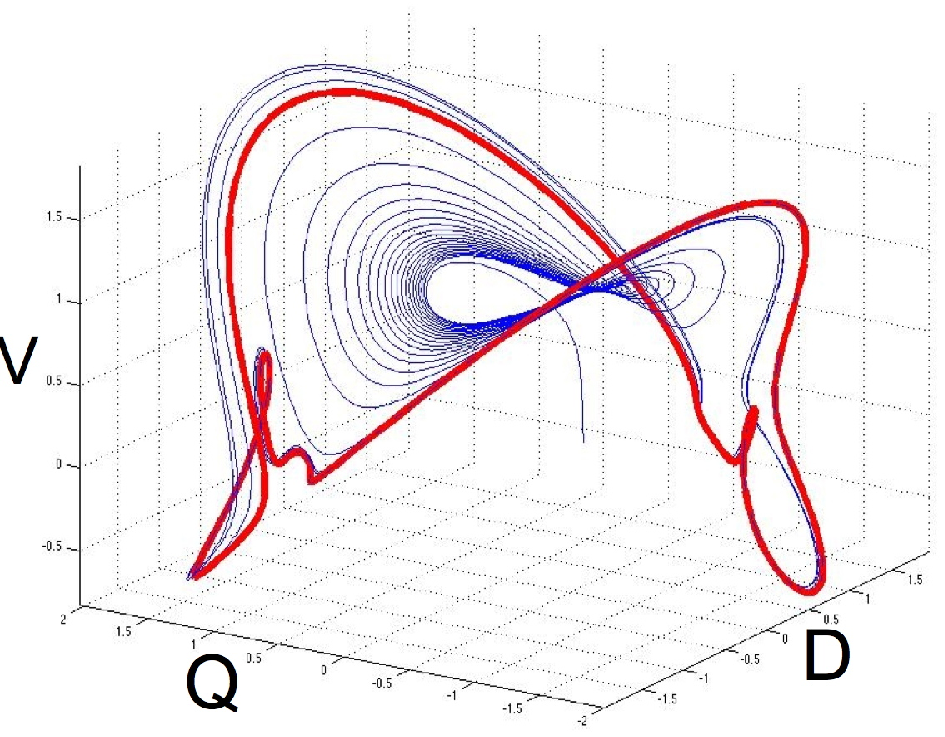}}
\vspace*{-3mm}
		\centerline{\includegraphics[width=0.85\columnwidth]{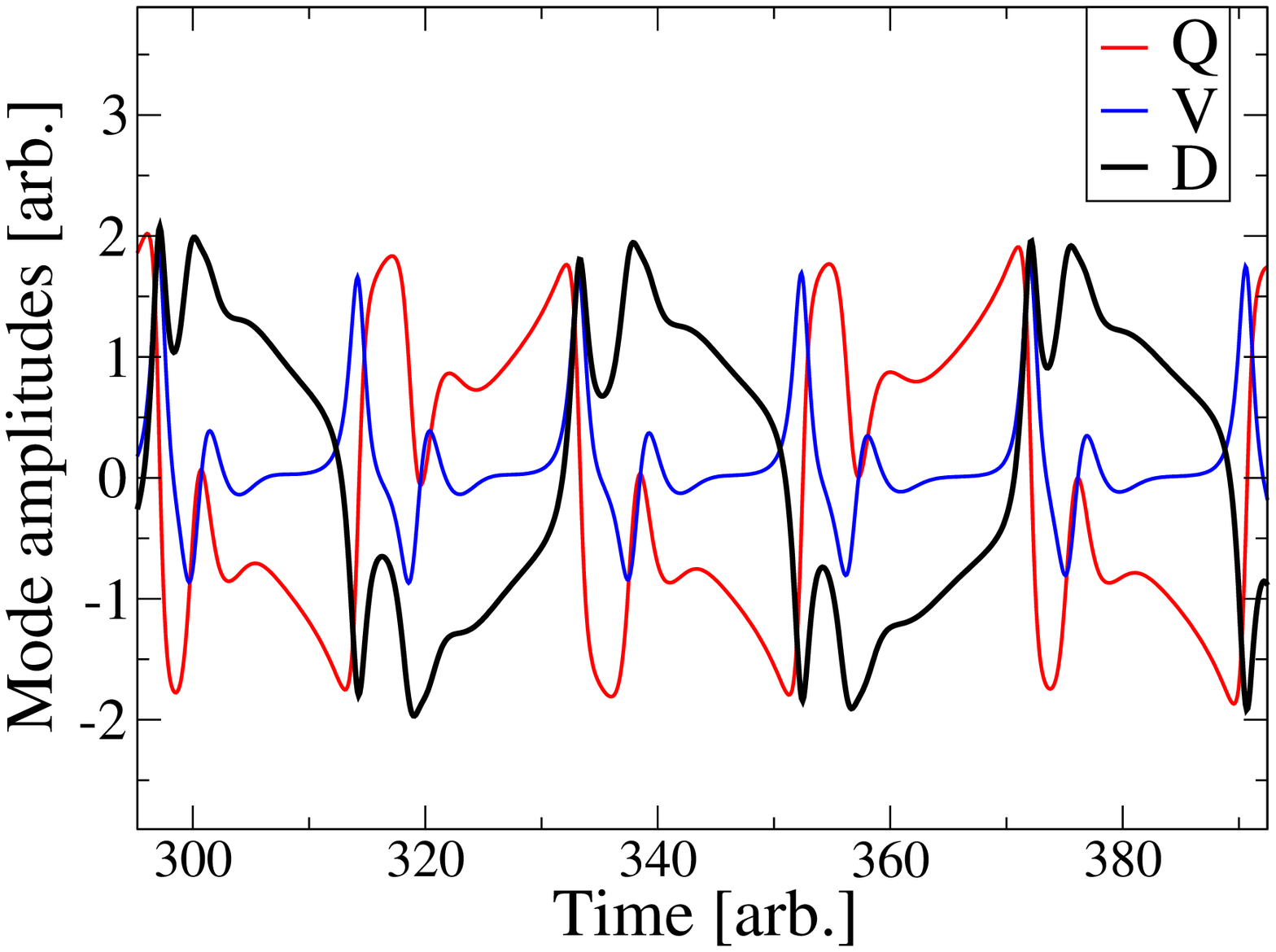}}
\caption{ Top: phase space for $\mu=0.101$, $\nu=0.1$ and
  $\Gamma=0.9$. Reversals are not chaotic anymore for these values. A
  periodic connection is achieved between the two fixed points. Bottom:
  Time evolution of the three amplitudes $D$, $Q$, $V$.}
\label{mod_mu0p101}
\end{figure}

As shown above, for $\Gamma=0.9$ and $\nu=0.1$, a decrease of $\mu$
leads to four well defined successive behaviors: a stationary state,
periodic then chaotic oscillations around this non-trivial fixed
point, and finally, chaotic reversals between the vicinity of the two
fixed points when $\mu<\mu_c$.  For further decrease of $\mu$, the
averaged waiting time between reversals strongly decreases and the
system finally reaches a regime of chaotic oscillations. Figure
\ref{mod_mu0p101} shows the evolution of the system as a function of
time and in phase space, for $\mu=0.101$, $\nu=0.1$ and $\Gamma=0.9$.
When $\mu$ finally reaches the value of $\nu$, we note that the system
becomes linearly stable at the origin $O$, since the real parts of
$\lambda_1$ and $\lambda_2$ vanish. However, this bifurcation appears
to be strongly subcritical: for initial conditions chosen inside the
basins of attraction of the non-trivial fixed points, the trajectories
stay inside this basin of attraction. The system stabilizes on the
trivial fixed point $O$ independently of initial conditions only for
$\mu<0.063$.

Inside this interesting regime of bistability, a series of transitions
is observed in the dynamics of our system as $\mu$ is decreased.
\begin{figure}[ht!]
	\centerline{
		\includegraphics[width=0.6\columnwidth]{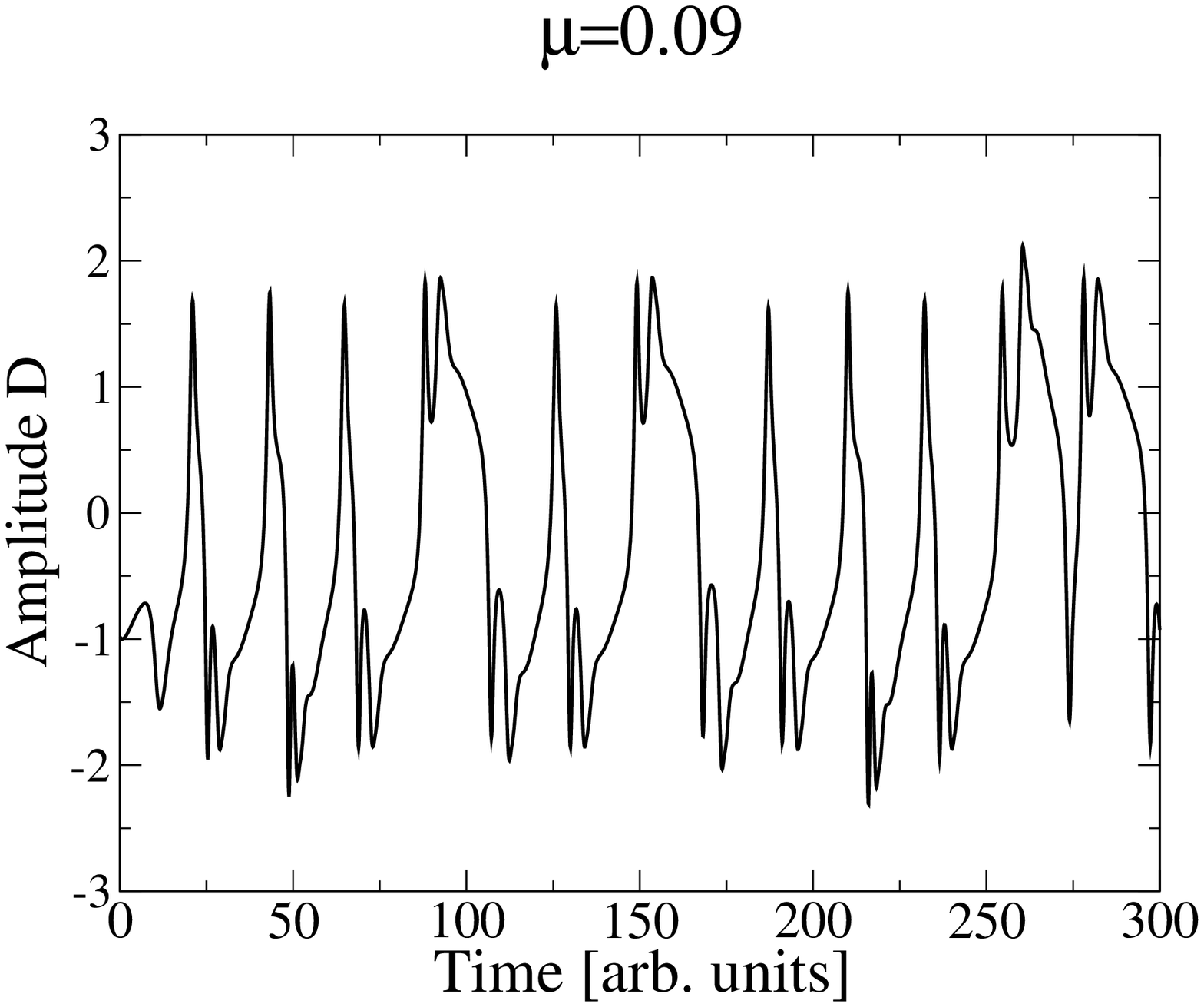}
                \includegraphics[width=0.6\columnwidth]{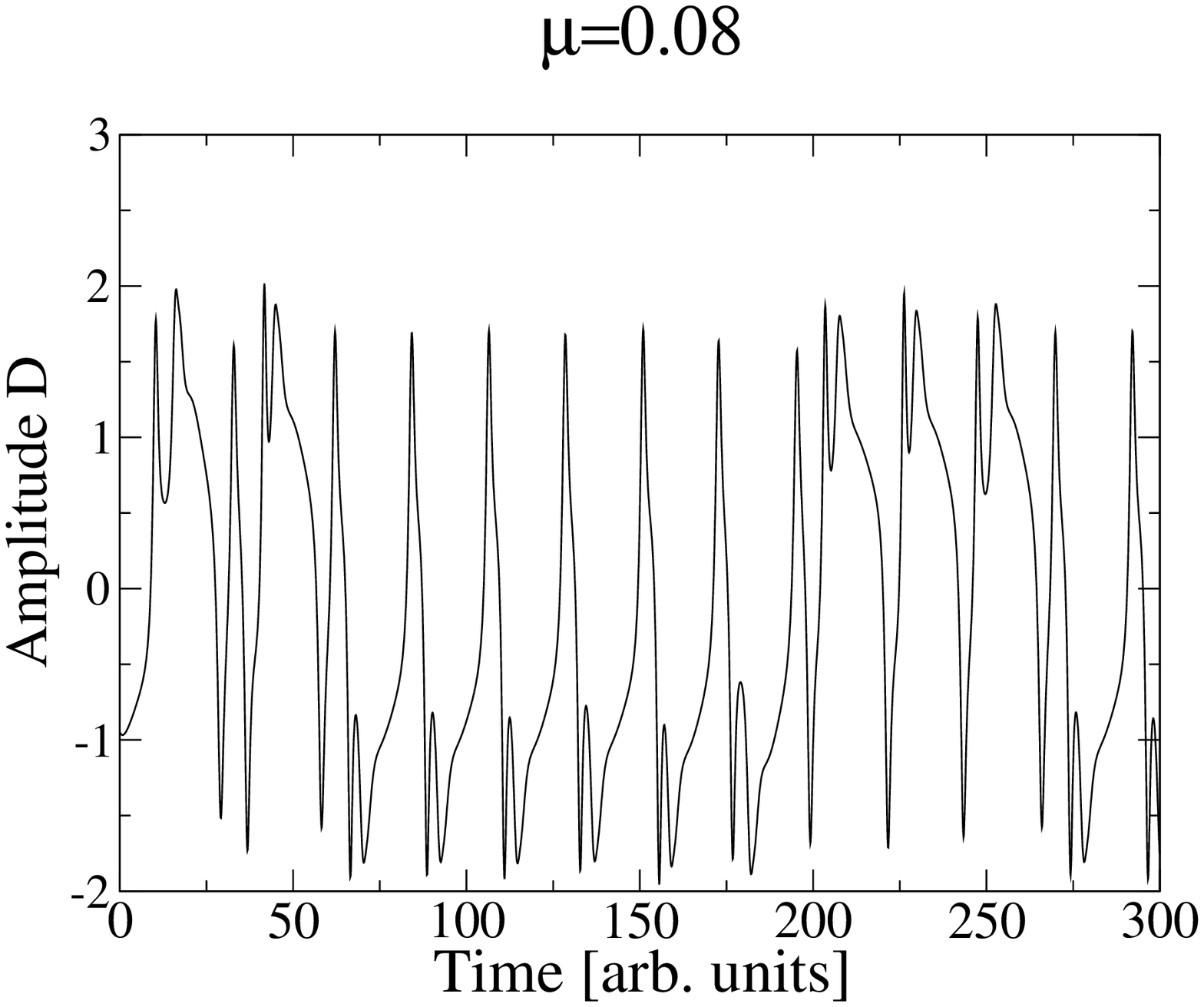}}
        \centerline{        \includegraphics[width=0.7\columnwidth]{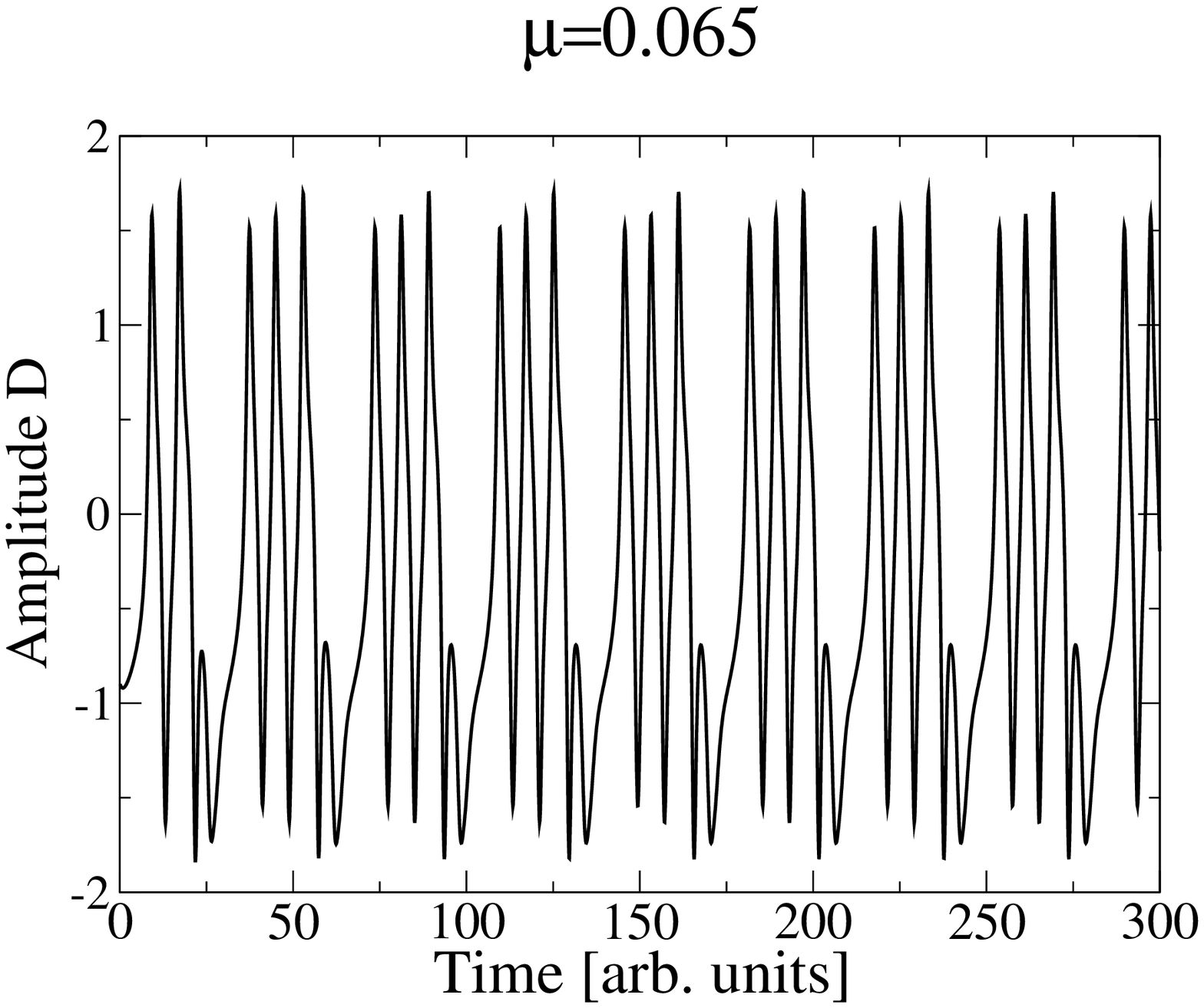}}
\caption{ Evolution of the amplitude $D$ for different values of
  $\mu$. Top, left: $\mu=0.09$. Top, right: $\mu=0.08$. Bottom:
  $\mu=0.065$. Note the asymmetric behavior of the dynamics in the
  latter case. However, the symmetric attractor can be obtained with
  different initial conditions. }
\label{mod_asym}
\end{figure}
As it can be seen in figure \ref{mod_asym}, for $0.065<\mu<0.1$, the
system explores the vicinity of one of the fixed point during a given
time and suddenly makes an oscillation around the trivial point
$O$. Although it stays on the basin of attraction of the same
attractor, the trajectory seems to briefly explore the opposite
state. During the time spent close to the fixed point, the behavior
can be slightly chaotic. We thus observe an intermittency between a
chaotic state around a fixed point and an unstable periodic orbit: as
time evolves, the system switches chaotically from the vicinity of the
fixed point to the unstable periodic orbit encircling the origin. As
expected, the mean time between these switches increase when $\mu$
decreases. For $\mu=0.065$, surprising asymmetric reversals can be
observed, despite the symmetry of equations.\\

\begin{figure}[ht!]
	\centerline{\includegraphics[width=0.85\columnwidth]{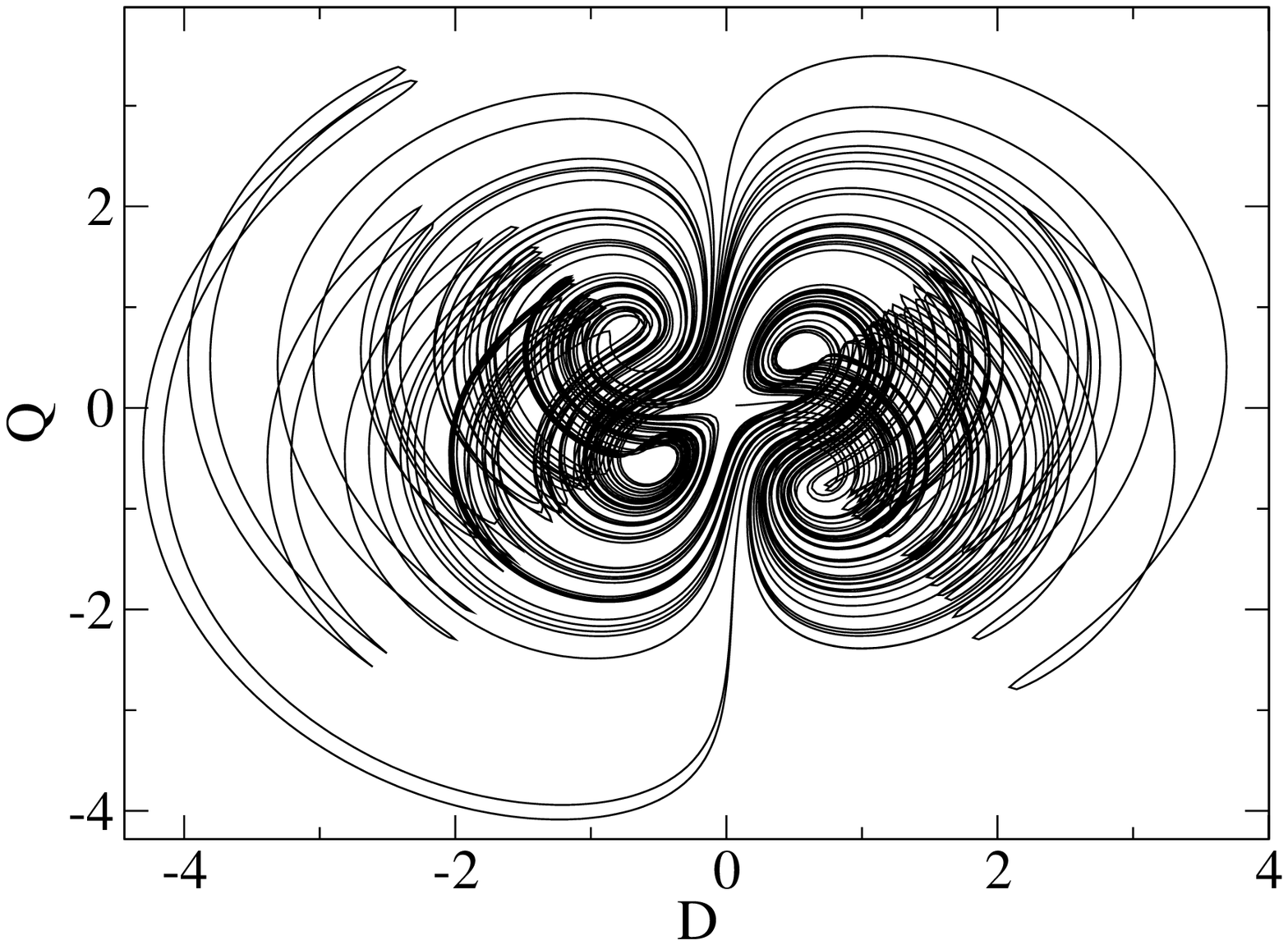}}
\vspace*{-3mm}
        \centerline{\includegraphics[width=0.85\columnwidth]{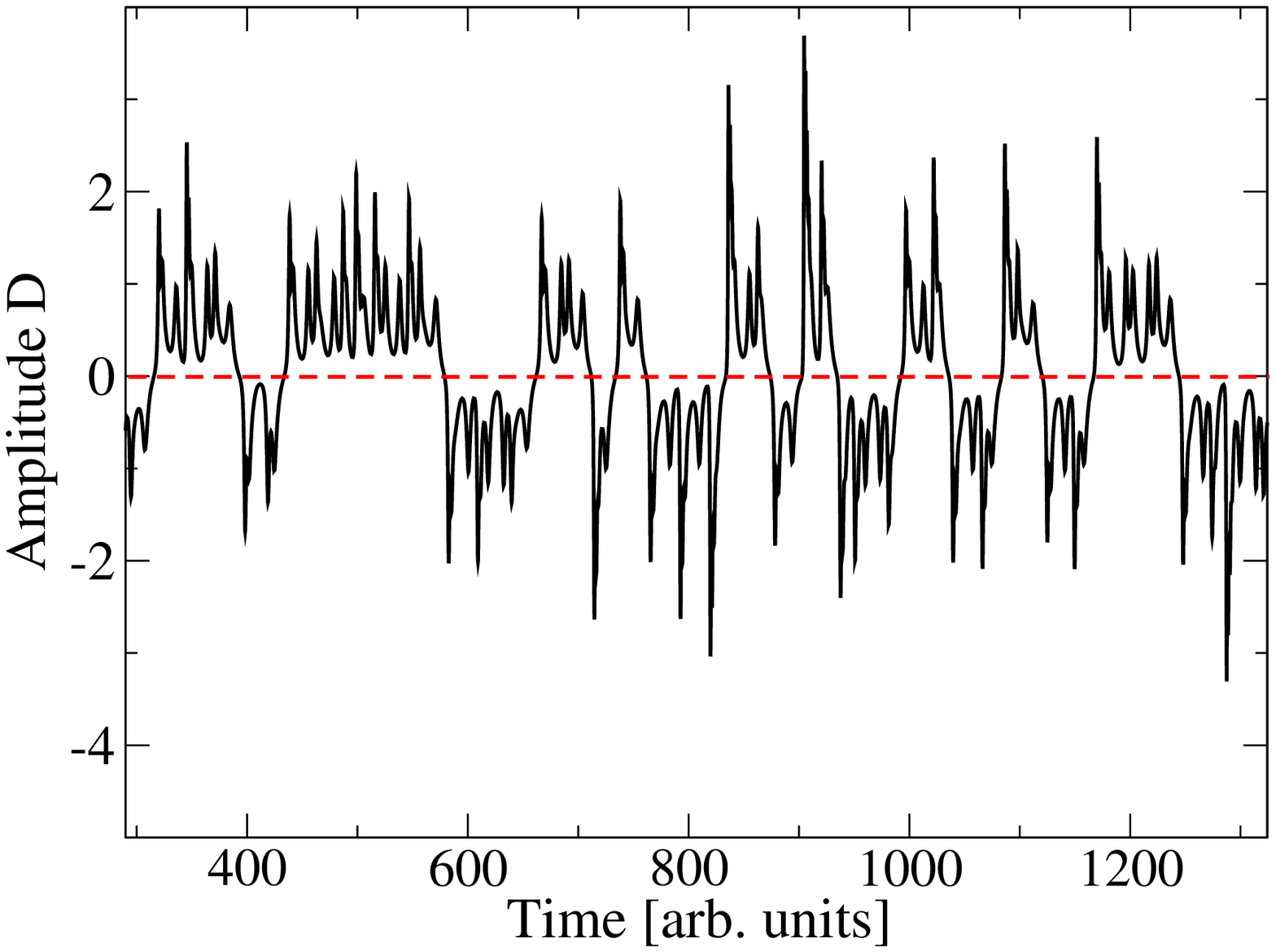}}   
\caption{ Reversals obtained for $\mu=\nu=0.35$ and $\Gamma=0.1$, when
  the four non-trivial fixed points are involved. Top: evolution in
  the phase space $(D,Q)$. Bottom: evolution of the amplitude $D$ as a
  function of time.}
\label{mod_4pt}
\end{figure}

The set of parameters chosen so far ensures the existence of only two
of the four non-trivial fixed points. We have seen that this is
already sufficient to generate several different dynamical regimes,
mainly related to reversals. One interest of this model is its ability
to describe very different regimes, without any needs of nonlinear
terms of higher degrees. This can be illustrated by choosing
parameters such that $\sqrt{\mu\nu}-\Gamma>0$, meaning that there are
now four non-trivial fixed points in addition to $O$. In this case,
the two additional solutions induce new dynamics. Figure \ref{mod_4pt}
shows the evolution of trajectories in the phase space $(D,Q)$ (top)
and as a function of time (bottom), for $\mu=\nu=0.35$ and
$\Gamma=0.1$.

\begin{figure}[ht!]
    \centerline{\includegraphics[width=0.7\columnwidth]{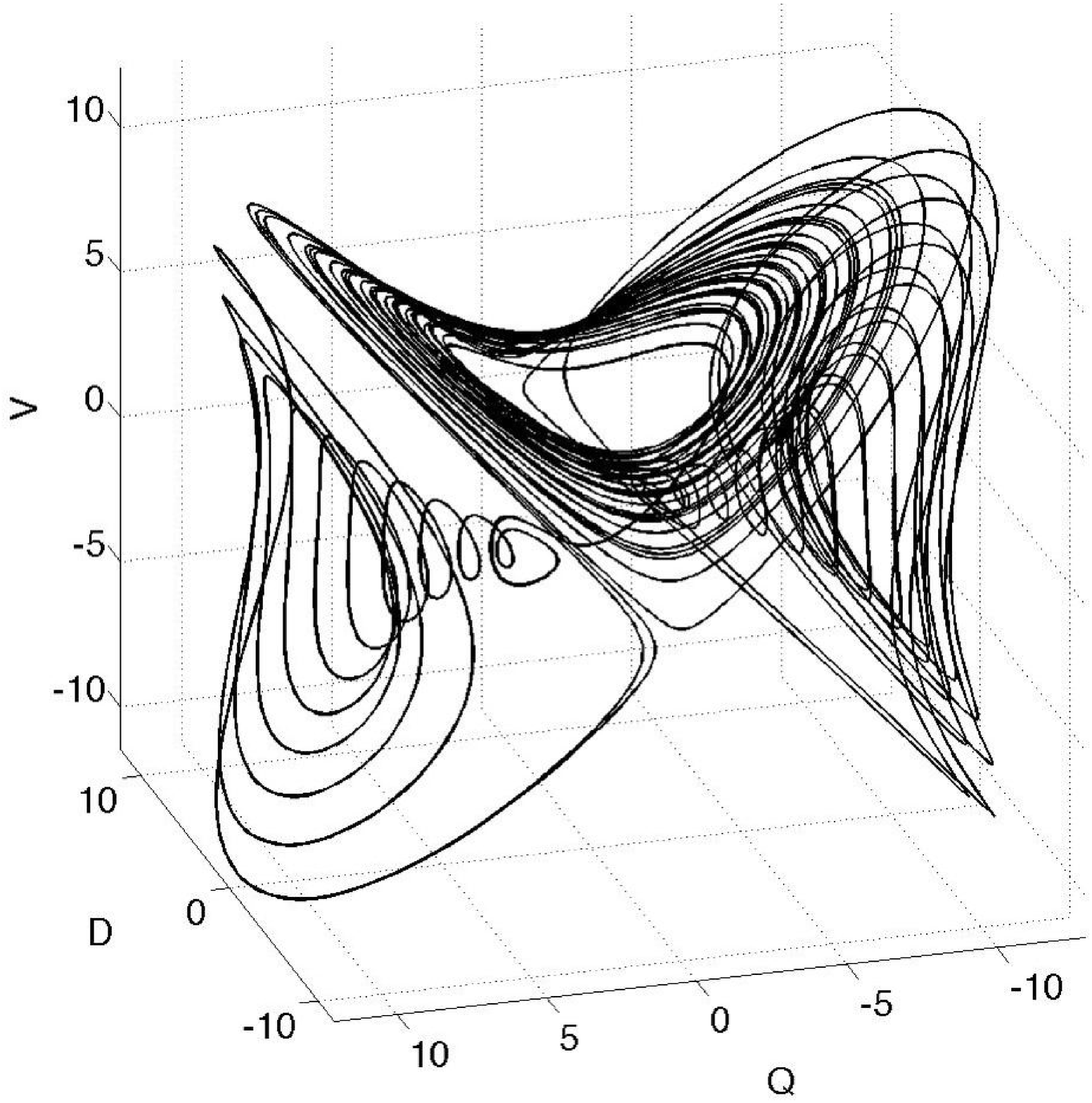}}
\vspace*{-3mm}
    \centerline{\includegraphics[width=0.8\columnwidth]{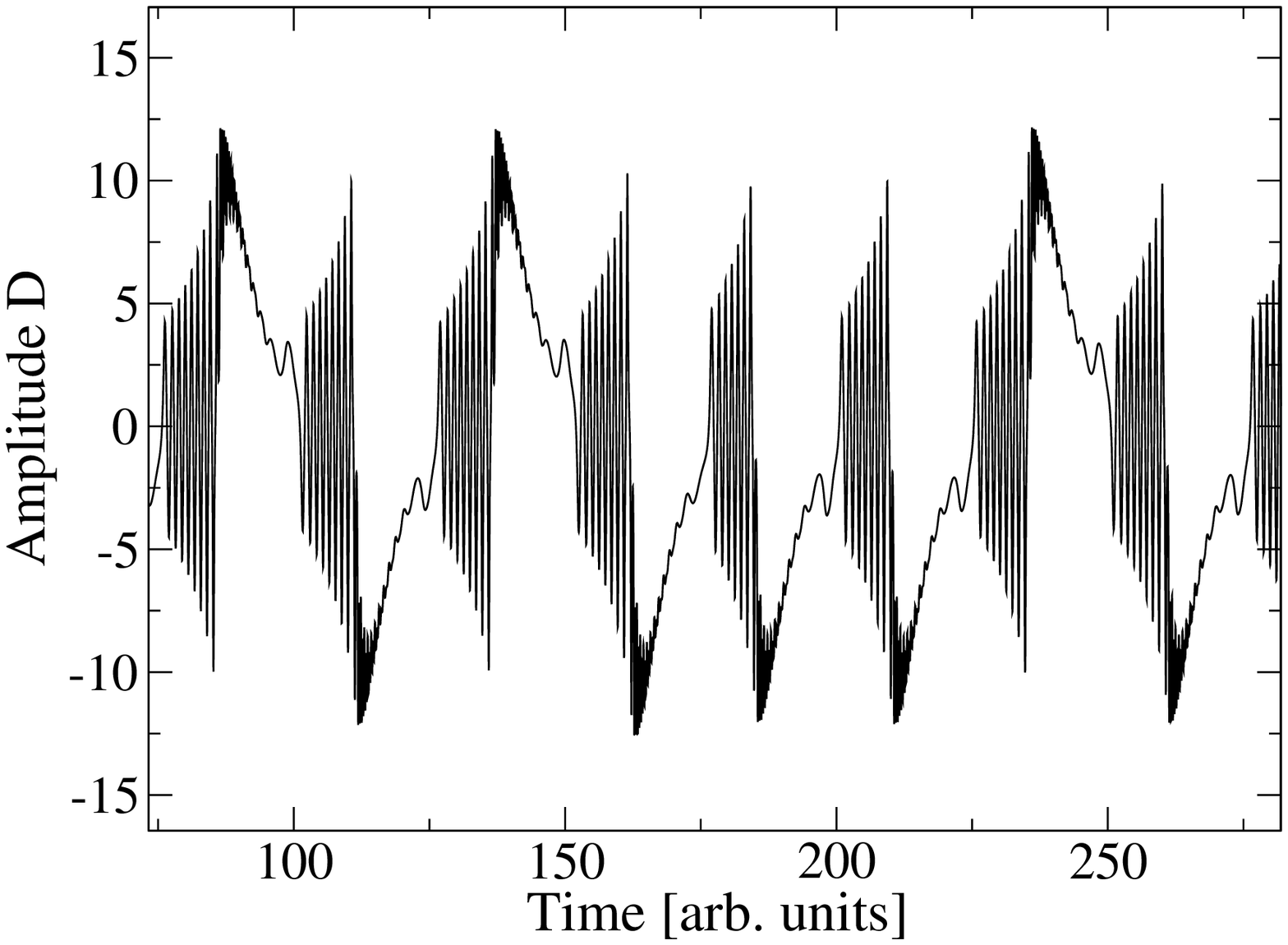}}   
\caption{ Evolution of the system for $\mu=0.3$, $\nu=0.2$ and
  $\Gamma=5$. Top: evolution in the phase space $(D,Q)$. Bottom:
  evolution of the amplitude $D$ as a function of time.}
\label{mod_shil}
\end{figure}

Finally, others exotic behaviors can be obtained. For instance, figure
\ref{mod_shil} shows the evolution of the system for $\mu=0.3$,
$\nu=0.2$ and $\Gamma=5$.
In this regime, an intermittency between two distinct behaviors is
observed. During a first phase, trajectories slowly escape from the
unstable periodic orbit surrounding the trivial point $O$. When the
amplitude of the oscillation is sufficiently large, it reaches the
basin of attraction of one of the two fixed points, and after some
transient wandering around this fixed point, the trajectory is
re-injected toward the center of the initial periodic orbit, and the
process repeat. Note that here, chaos is essentially due the apparent
random choice of the fixed point involved at the end of the
oscillations around $O$.\\

Compared to previous similar models, this simple dynamical system,
involving only $3$ quadratically coupled equations, generates several
different dynamical behaviors, depending on the values of the free
parameters $\mu$, $\nu$ and $\Gamma$. It would be interesting to see
at which point the different characteristics described here are
robust. For instance, one could add multiplicative or additive noise
to equations $(\ref{mod_ampD}-\ref{mod_ampV})$, or take into account
terms of higher degrees to couple the three amplitudes. The present
form of the model (without noise and limited to quadratic terms)
presents of course the advantage to remain extremely simple, and yet
generating complex dynamical regimes.

\section{Physical interpretation and comparison with geodynamo reversals}
\label{sec:3}
The perspective of a description of the reversals of the Earth's
magnetic field in the framework of a low dimensional model is very
tempting. Indeed, a low dimensional description provides a simple
vision of the phenomena, and allows to make interesting predictions
concerning the shape and the statistical properties of the geomagnetic
reversals.  In fluids dynamos, the small ratio between kinematic
viscosity and magnetic resistivity (small magnetic Prandtl number),
together with the proximity to the dynamo threshold, strongly reduce
the degrees of freedom of the system, and low-dimensional dynamics can
be expected.  For this reason, this approach has been followed by
several studies in the past \cite{Noziere78}, \cite{Rikitake58},
\cite{Melbourne01}, \cite{Hoyng01}, \cite{PFD}. Unfortunately, despite
their chaotic behaviors, low dimensional deterministic models
generally yield too simple dynamics, and fail to realistically
reproduce observational data.  For instance, Lorenz and Rikitake
models exhibit typical growing oscillations between reversals,
characteristics of their low-dimensionality, but absent from
geomagnetic measurements.

\begin{figure*}[ht!]
	\centerline{
                    \includegraphics[width=0.8\columnwidth]{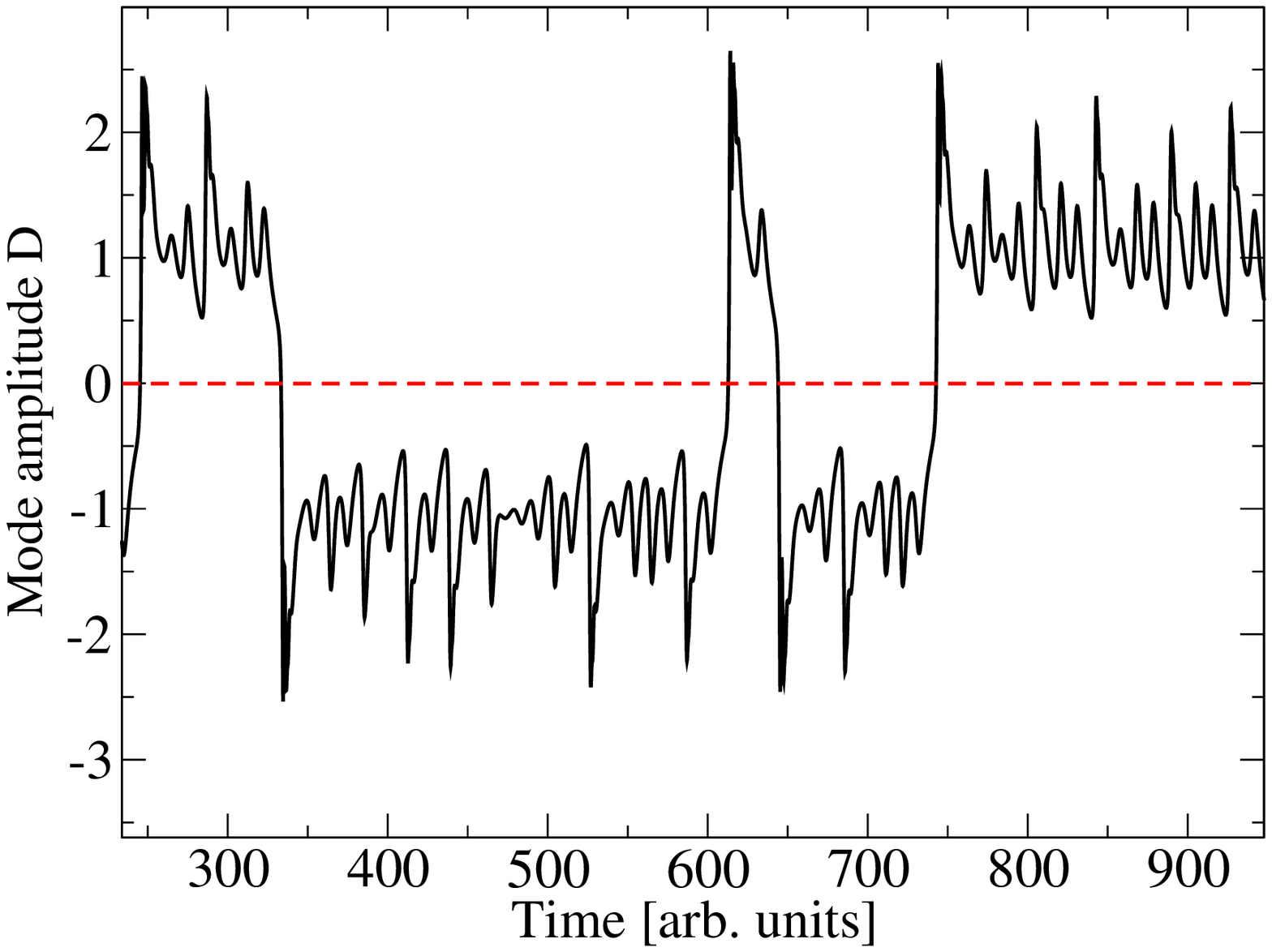}
                    \includegraphics[width=0.85\columnwidth]{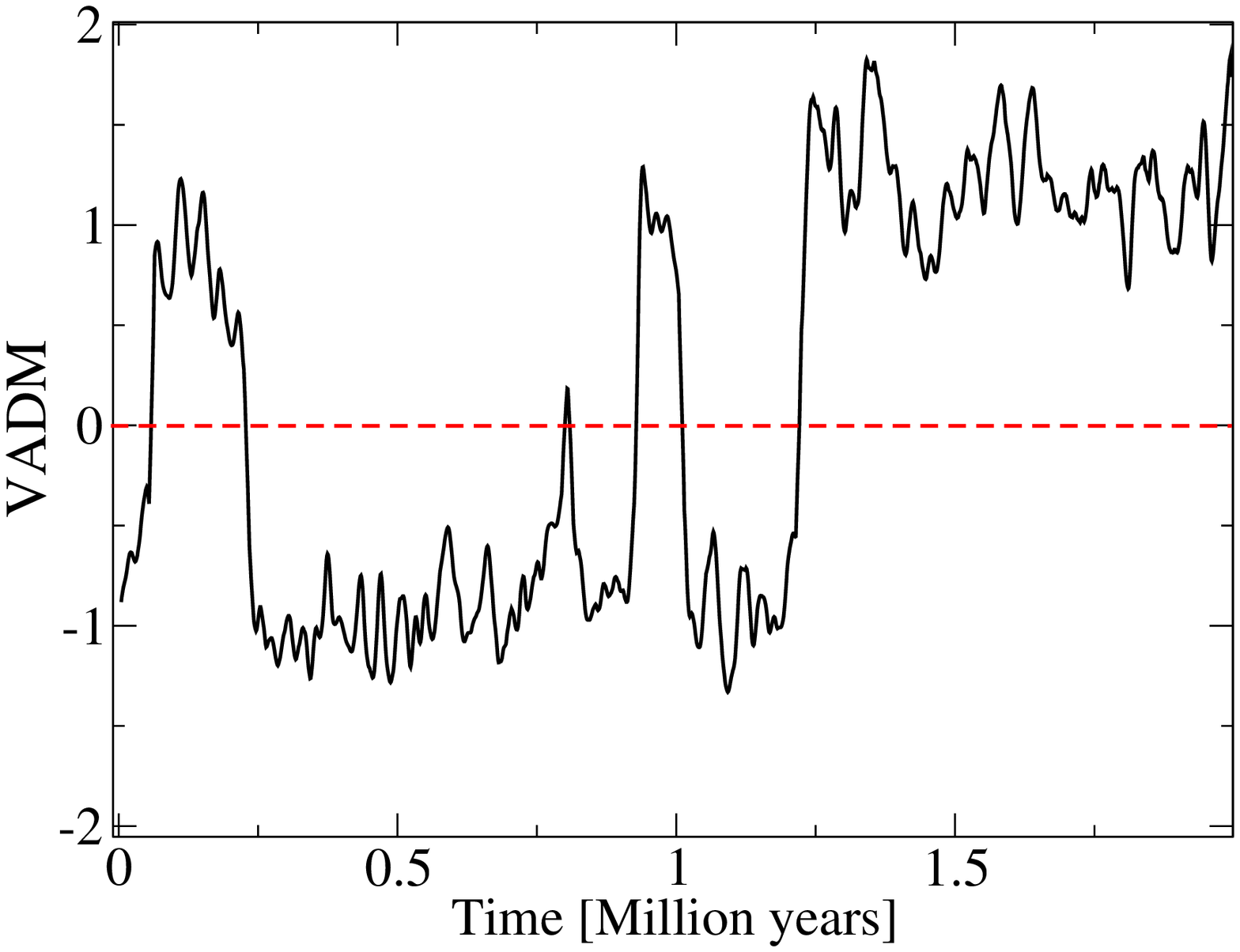}}
        \centerline{            \includegraphics[width=0.85\columnwidth]{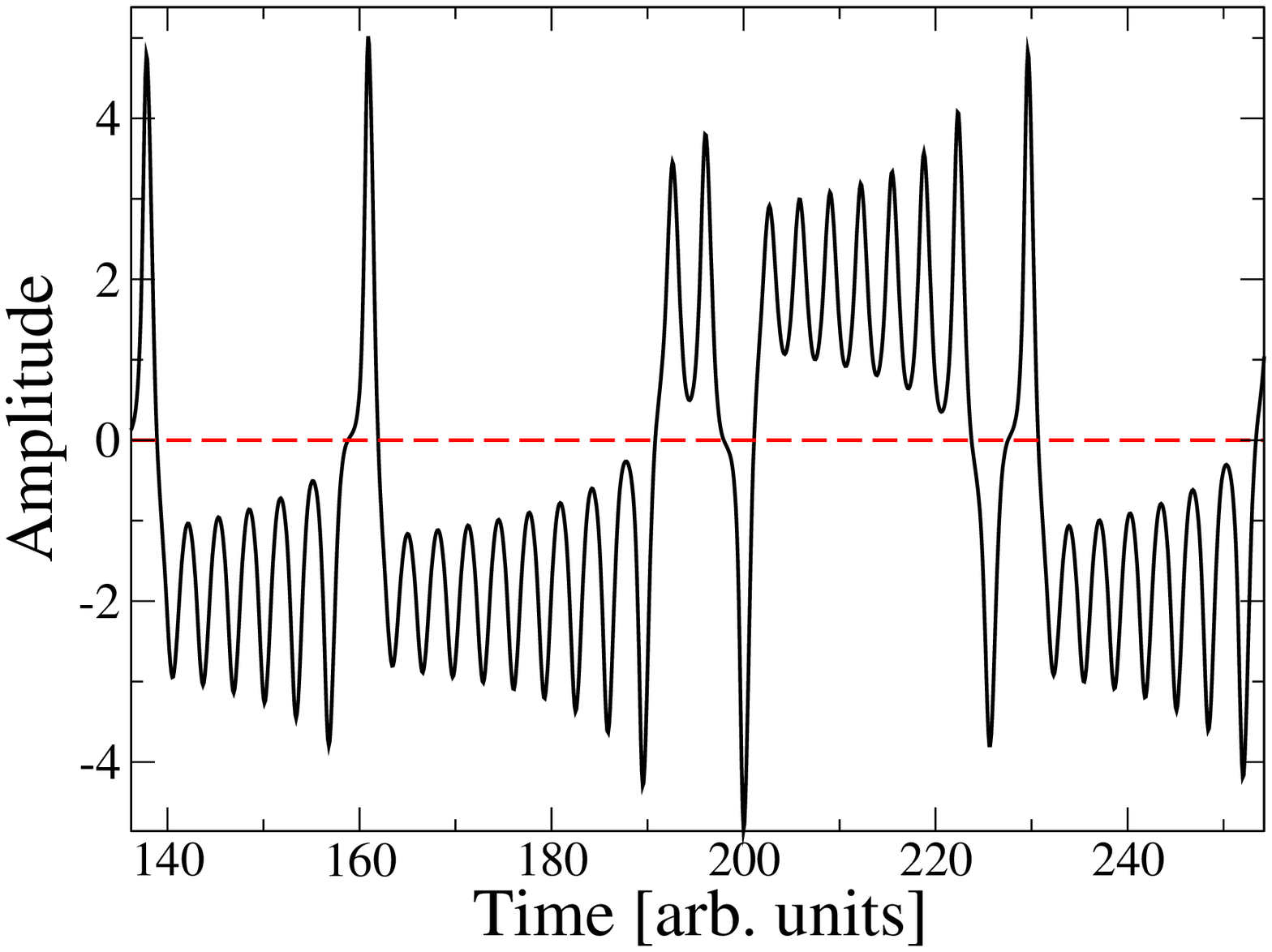}}
 \caption{ Comparison between time series obtained from paleomagnetic
   data (top,right)(SINT2000 data \cite{Valet05}) and from the model
   $(D,Q,V)$ with $\mu=0.119$, $\nu=0.1$ and $\Gamma=0.1$
   (top,left). A typical time series obtained from the Lorenz model is
   shown at the bottom. }
\label{mod_mod}
\end{figure*}

As shown in the previous sections, reversals of our simple model are
generated by crisis-induced intermittency, a mechanism different from
the one creating chaotic reversals in classical models. We will show
in this section that the model $(\ref{mod_ampD}-\ref{mod_ampV})$ leads
to a better agreement with geodynamo reversals than Lorenz and
Rikitake models. Moreover, using symmetry arguments, it is possible to
give a simple physical interpretation to the three amplitudes.\\

The magnetic field of the Earth is mainly dominated by its dipolar
component. Figure \ref{mod_mod}(top-right) shows the evolution of the
Earth dipole moment during the past two millions years
\cite{Valet05}. Although its amplitude is not known during reversals,
it has been suggested that the quadrupolar mode could also play an
important role during reversals \cite{McFadden91},
\cite{Glatzmaier95}. In the model $(\ref{mod_ampD}-\ref{mod_ampV})$,
amplitudes $D$ and $Q$ can be regarded as dipolar and quadrupolar
components of the magnetic field. Note that the system is invariant
with respect to the symmetry $(D\rightarrow -D$, $Q\rightarrow -Q)$,
in agreement with the symmetry $(B\rightarrow -B)$ of the magnetic
field $B$ in the equations of magnetohydrodynamics. Another important
symmetry in the Earth core is the mirror symmetry with respect to the
equatorial plane. Under this symmetry, the dipole is unchanged whereas
the quadrupole is transformed in its opposite. This implies that the
dipole and quadrupole can not be linearly coupled if the flow in the
Earth core is mirror symmetric. By noting that the amplitude $V$ is
coupling $D$ and $Q$ in equations $(\ref{mod_ampD}-\ref{mod_ampV})$,
we can therefore interpret $V$ as a velocity mode (externally forced
by $\Gamma\ne 0$) which breaks the mirror symmetry of the Earth core
flow. The term $QD$ in equation \ref{mod_ampV} thus stands for the
Lorentz force (quadratic in the magnetic field $B$). The system of
equations $(\ref{mod_ampD}-\ref{mod_ampV})$ therefore models the
interaction between dipolar and quadrupolar magnetic modes coupled by
a mirror-antisymmetric velocity mode, and can be directly compared to
geomagnetic data.

This interpretation of the model is in fact very close to a recent
model proposed by Petrelis and Fauve, also relying on a
dipole-quadrupole interaction \cite{Petrelis09}, and aiming to model
the dynamics observed in the VKS (Von Karman Sodium) experiment, a
laboratory turbulent fluid dynamo \cite{Monchaux07}. The VKS
experiment exhibits chaotic and periodic reversals of the polarity of
the magnetic field, resulting from a surprisingly low-dimensional
dynamics \cite{Berhanu07}, and very similar to the predictions of the
Petrelis-Fauve model. In this model however, chaos is generated by the
addition of a stochastic noise in the equations (modeling the
turbulent fluctuations), and the velocity mode $V$ is not explicitly
taken into account. Nevertheless, it is interesting to note that most
of the regimes of the VKS experiment can also be reproduced with our
simple model, despite its deterministic dynamics
\cite{GissingerThesis}. This underlines the fact that at small $Pm$,
the characteristics of the turbulent flow and the details of the
configuration (such as boundary conditions) can be ignored as far as
the dynamical regimes of the magnetic field are considered.\\

\begin{figure*}[ht!]
	\centerline{
          \includegraphics[width=0.83\columnwidth]{eps_WTD.eps}
          \includegraphics[width=0.85\columnwidth]{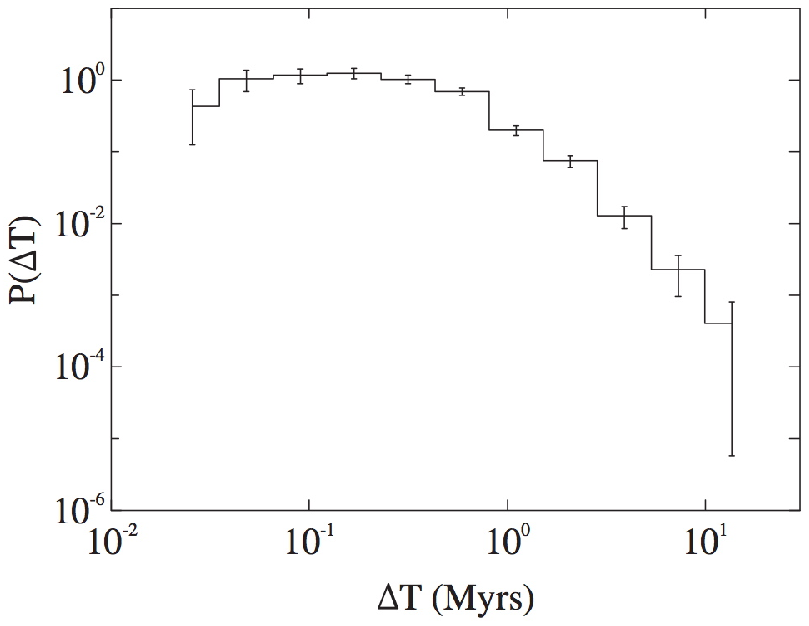}}
\caption{ Left: Distribution of the waiting time durations (WTD) for
  the model. We observe large oscillations at small time corresponding
  to the quantification of the unstable periodic orbits, and a
  poissonian behavior at large time. The red curve shows an
  exponential fit. Right: Probability distribution of waiting times
  between reversals for the geomagnetic field, showing an
  approximative poissonian law (from CK95 database \cite{CK95}).}
\label{mod_WTD}
\end{figure*}

Kono has shown in 1987 that the Rikitake model was unable to reproduce
the Probability Density Function (PDF) of waiting times between
reversals, or the PDF of the amplitude of the dipolar field
\cite{Kono87}.  In general, distributions of the averaged waiting time
between reversals are assumed to be poissonian for the geomagnetic
field . It is however difficult to conclude, since only a small number
of reversals are available \cite{Carbone06}. The Rikitake model
clearly fails to describe a poissonian process: depending on the
parameters, the model predicts either a cutoff forbidding long events,
or the apparition of too many long events. We have computed the
distribution of the waiting times for $\mu=0.119$, $\nu=0.1$ and
$\Gamma=0.9$, and figure \ref{mod_WTD} shows a comparison to the
paleomagnetic data, obtained from the CK95 database \cite{CK95}. One
can see that the very short times are characterized by a
quantification. This obviously comes from the minimal time it takes to
the system to follow at least one periodic orbit around a given fixed
point before reversing. This behavior is typical of models with
low-dimensional chaos. For larger waiting times however, a poissonian
law is obtained, with a distribution
\begin{equation}
P(\tau)={1\over<\tau>}e^{-{\tau\over<\tau>}}
\end{equation}
Our model is thus able to reproduce the poissonian behavior observed
for the Earth magnetic field, despite the small number of degrees of
freedom. Because the quantification at small time is a strong property
of deterministic chaos, it would be interesting to search for such a
behavior in geomagnetic observations. Note however that this type of
behavior is easily destroyed in the presence of noise, which cannot be
avoided in the Earth core due to the turbulent fluctuations of the
flow.\\

Similarly, there is a strong disagreement between observations and
classical models on the distribution of the intensity of the
dipole. The geodynamo exhibits two symmetric maxima around a minimum
at zero, but Rikitake and Lorenz models statistically spend too much
time close to zero, giving a different distribution. 
\begin{figure*}[ht!]
	\centerline{
		\includegraphics[width=0.85\columnwidth]{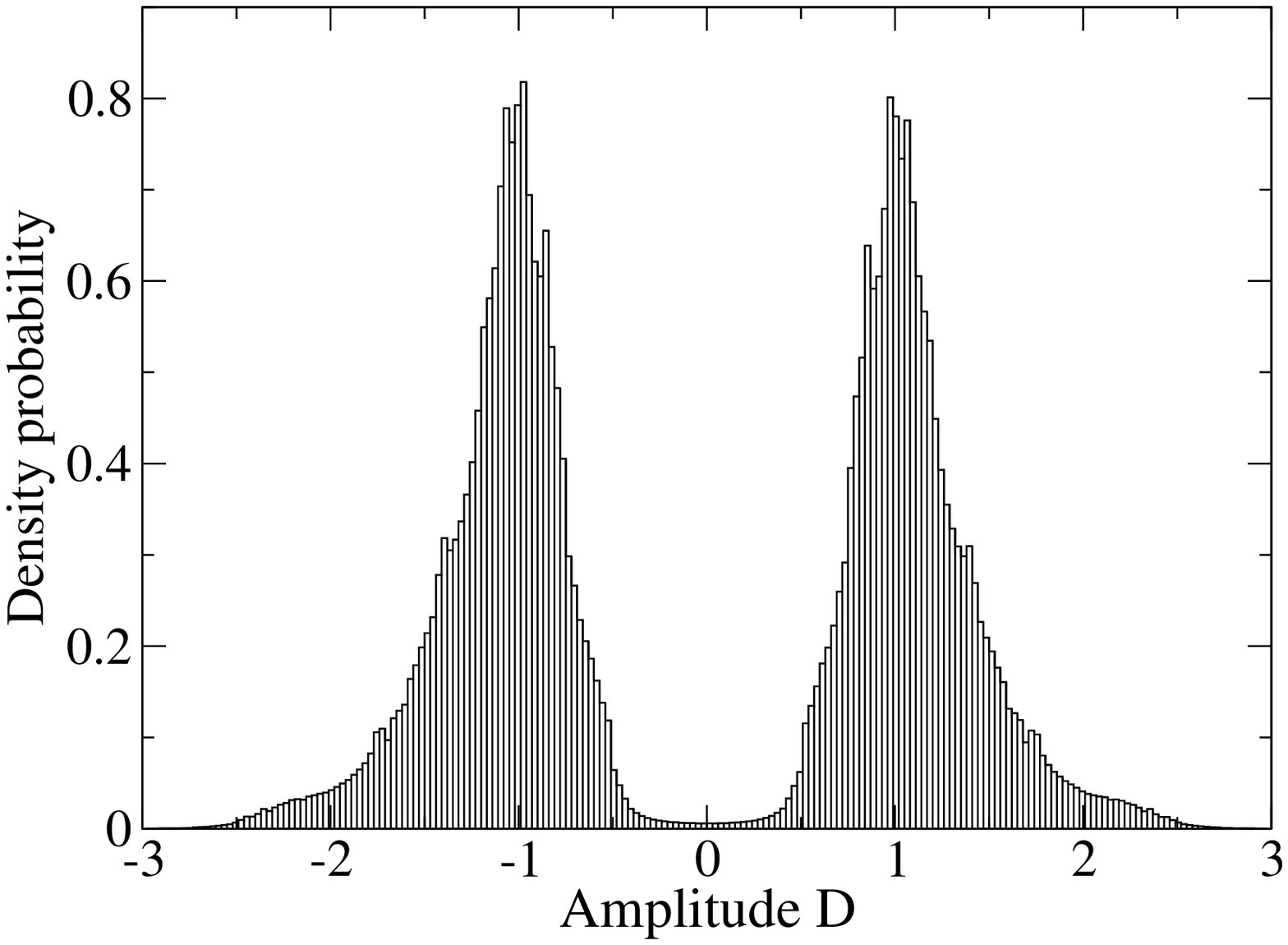}
                \includegraphics[width=0.85\columnwidth]{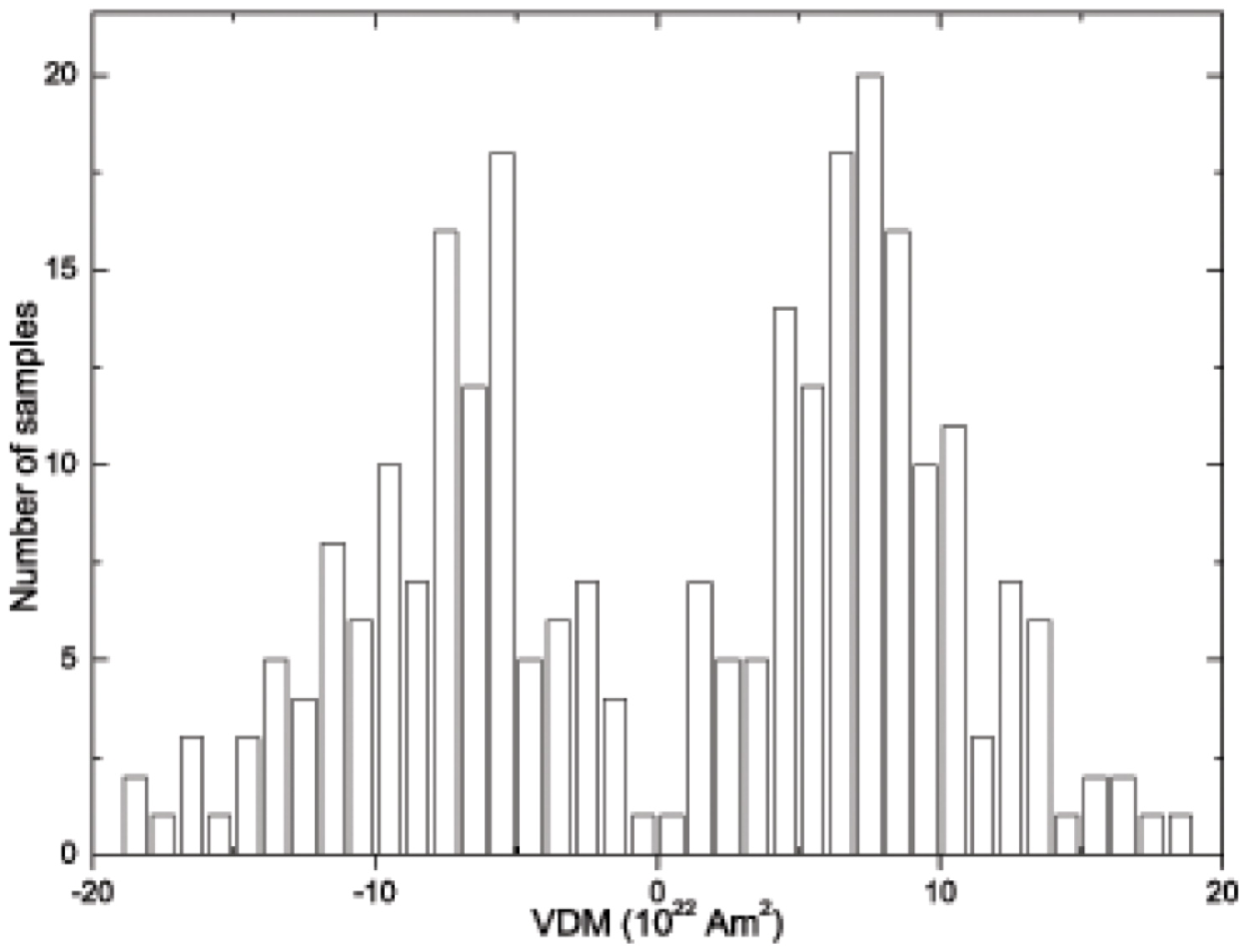}} 
\caption{ Left: distribution of probability of amplitudes $D$. Right:
  bimodal probability distribution of the intensity of the Earth
  dipolar field. }
\label{mod_pdf}
\end{figure*}
A study of the distribution of the amplitude $D$ in our model during
reversals shows a bimodal distribution with a very low probability
at the origin, in agreement with geomagnetic observations. This
success of the model is mainly due the the type of chaos generated in
the model, which produces a time scale separation between the duration
of a reversal (related to some unstable periodic orbit) and the
waiting time durations (controlled by the distance from the crisis
bifurcation point).\\

\begin{figure}[ht!]
		\centerline{\includegraphics[width=0.85\columnwidth]{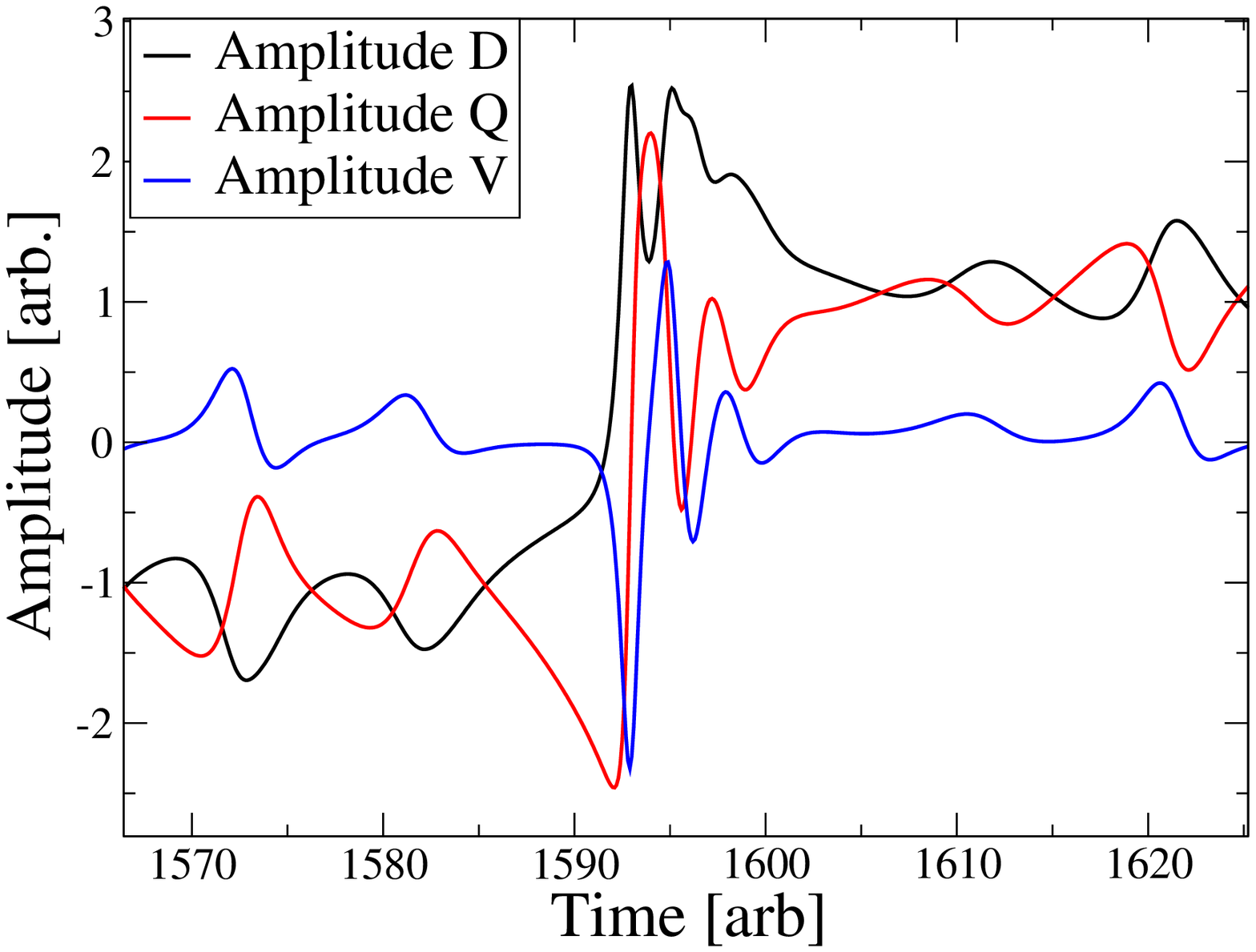}}
\caption{ Evolution of the three components of the model during a
  reversal, for $\mu=0.119$, $\nu=0.1$ and $\Gamma=0.9$. Note the
  rapid time scale related to the reversal event, compared to the rest
  of the dynamics. When the dipole D goes to zero during the reversal,
  the magnetic energy is transferred to the quadrupolar component of
  the magnetic field. After the reversal, the dipole systematically
  overshoots its mean value.}
\label{mod_attract2}
\end{figure}

From a general point of view, we see that this new model allows to
reproduce some of the main characteristics of the geodynamo, like the
poissonian process or the distribution of the intensity of the field.
As it appears in figure \ref{mod_mod}, this resemblance between the
two systems is also apparent from a simple examination of the time
series: during a stable polarity period, both the model and the Earth
magnetic field show some chaotic oscillations, with sometimes bigger
events called excursions, during which the dipole seems to reverse but
finally comes back to its initial value. Moreover, this model
reproduces the tendency of the Earth dipolar field to systematically
overshoot its quasi-stationary value just after each reversal, and
also the time scale separation between the reversals and the stable
polarity states. This model also makes some interesting predictions
concerning the geodynamo. For instance, when the dipole $D$ goes to
zero during a reversal, this corresponds to a strong increase of the
quadrupolar component $Q$ in the model (see figure
\ref{mod_attract2}).  Note that because of the deterministic nature of
the model, the reversals always exhibit these features. This predicts
that during a reversal, the magnetic field of the Earth would change
shape (by transferring the magnetic energy from the dipolar to the
quadrupolar component), rather than simply vanishing.

In addition, this model relies on the existence of a velocity mode $V$
breaking the mirror symmetry of the Earth core flow. Several studies
seem to confirm this role of an equatorially antisymmetric velocity
mode in the geodynamo reversals. First, it has recently been observed
that the ends of superchrons (large period of time without geomagnetic
reversals) correspond to major flood basalt eruptions due to large
thermal plumes ascending through the mantle \cite{Courtillot07}. In
the framework of our model, these plumes correspond to an increase in
the symmetry breaking of the flow ($\Gamma$ is increased). In
agreement with this observation, it has been shown in global numerical
simulations that taking an heterogeneous heat flux (rather than
homogeneous) at the core-mantle boundary of the Earth strongly
influences the frequency of magnetic field reversals
\cite{Glatzmaier99}. Again, this corresponds in the model to change
the forcing of the antisymmetric velocity mode $V$.

\section{Conclusion}
\label{conclu}
In this paper, we have presented a new deterministic model for
reversals. In this model, chaotic reversals are generated through a
boundary crisis between two symmetric chaotic attractors. Compared to
previous deterministic models, this mechanism produces a relatively
complex behavior in a given polarity, while reversals exhibit a very
robust shape. In addition, we have shown that several very different
dynamical regimes can be obtained, depending on the values of the free
parameters of the system.\\

This model has been compared to reversals of the Earth magnetic field,
and a good agreement has been obtained. Although geomagnetic reversals
are certainly more complicated than a simple interaction between three
modes, it is interesting to obtain a simple way to describe this
complex phenomenon. The strong similarities between our model and
paleomagnetic observations suggest an interpretation of geomagnetic
reversals as resulting from an interaction between the dipolar and the
quadrupolar components of the magnetic field. In addition, our model
predicts that a velocity mode could play the role of the amplitude
$V$, by coupling the two magnetic components dipole and quadrupole. It
would be interesting to study the different possibilities for such a
flow, corresponding to a breaking of the mirror symmetry of the large
scale flow in the Earth core.  Because of the fast rotation of the
Earth, the geostrophic equilibrium makes difficult a spontaneous
breaking of this symmetry. However, the presence of the core,
decoupling the north and the south hemispheres inside the tangent
cylinder, makes possible the existence of large scale antisymmetric
velocity modes: for instance, an heterogeneous heat flux at the
core-mantle boundary may force the large scale velocity mode $V$.\\

Several physical systems, different from the dynamo, display chaotic
reversals of an observable between two symmetric states. In general,
the interpretation in terms of a few coupled equations generating
deterministic chaos is easily ruled out. Indeed, most of the models
predict a behavior far too simple between reversals, like the growing
periodic oscillations of the Lorenz or Rikitake equations. It is
possible to attenuate this behavior by adding noise to the system,
taking into account higher order terms, or increasing the dimension of
the model. The new model presented in this article shows that it is in
fact possible to obtain a more complex behavior between reversals,
closer to stochastic models predictions, with only three coupled
equations limited to quadratic orders. This is an important
observation, because it opens the perspective to understand the
dynamics of complex physical systems using very simple mechanisms,
mainly relying on symmetry arguments.\\

The author would like to sincerely thank Stephan Fauve, Francois
Petrelis and Emmanuel Dormy for their very inspiring discussion and
constructive comments. This work was supported by the NSF under grant
AST-0607472 and by the NSF Center for Magnetic Self-Organization
under grant PHY-0821899.\\



\begin{thebibliography}{5}

\bibitem{Wiggins90}{\sc S. Wiggins}, {\em Introduction to applied nonlinear
  dynamical systems and Chaos}, Springer, Berlin (1990)
\bibitem{Ott93}{\sc E. ott}, {\em Chaos in dynamical systems}, Cambridge Univ. Press, Cambridge U.K. (1993)
\bibitem{GuckHolmes}{\sc J. Guckenheimer and P. Holmes}, {\em Nonlinear
  oscillations, dynamical systems, and bifurcations of vector fields},
  Applied Mathematical Sciences, New York: Springer (1982)
\bibitem{Lorenz63}{\sc E.N. Lorenz}, {\em Deterministic non-periodic
  flow}, J. Atmos. Sci. 20, 130-141 (1963)
\bibitem{Rossler76} {\sc O.E. Rossler}, Phys. Rev. Lett. {\bf 57A}, 397 (1976)
\bibitem{Rikitake58} {\sc T. Rikitake}, Math. Proc. Cambridge Philos. Soc., {\bf 54} 89-105 (1958)
\bibitem{Sprott94} {\sc J.C. Sprott}, Phys. Rev. E, {\bf 50}, 647 (1994)
\bibitem{Gissinger10} {\sc C. Gissinger, E. Dormy, S. Fauve}, Euro. Phys. Lett. {\bf 90} 49001 (2010)
\bibitem{Noziere78} {\sc P. Nozi\`eres}, Phys. Earth Planet. Inter., {\bf 17}, 55 (1978)
\bibitem{Hughes90} {\sc D. Hughes and M.R.E. Proctor}, Nonlinearity, {\bf 3} (1990) 127.
\bibitem{Smith98} {\sc P. Smith}, Explaining chaos, Cambridge University Press, Cambridge UK(1998)
\bibitem{Grebogi82} {\sc C Grebogi, E Ott, J.A. Yorke}, Phys. Rev. Lett. {\bf 48}, 1507-1510,(1982)
\bibitem{Grebogi87} {\sc C Grebogi, E Ott, F Romeira, J.A. Yorke}, Phys. Rev. A,{\bf 36}, 5365-5380 (1987)
\bibitem{Melbourne01} {\sc I. Melbourne, M.R.E. Proctor and A.M. Rucklidge}, Dynamo and Dynamics, A Mathematical Challenge, edited by P. Chossat et al. (Kluwer Academic Publishers) 2001, p. 363
\bibitem{Hoyng01} {\sc P. Hoyng, M. A. J. H. Ossendrijver, D. Schmidt} ,
  Geophys. Astroph. Fluid Dyn. {\bf 94} (2001) 263-314.
\bibitem{PFD} {\sc F. Petrelis, S. Fauve, E. Dormy, J.P. Valet}, Phys. Rev. Lett., {\bf 102} (2009) 144503.
\bibitem{Valet05} {\sc J.P. Valet, L. Meynadier and Y. Guyodo Y}, Nature, {\bf 435}, 802 (2005)
\bibitem{McFadden91} {\sc P. L. McFadden et al.}, J. Geophys. Res. {\bf 96}, 3923 (1991) 
\bibitem{Glatzmaier95} {\sc G. Glatzmaier and P. Roberts}, Phys. Earth Plan. int., {\bf 91}, 63-75 (1995) 
\bibitem{Petrelis09} {\sc F. Petrelis and S. Fauve}, J. Phys. Condens. Matter {\bf 20}, 494203 (2008).
\bibitem{Monchaux07} {\sc R. Monchaux et al.},  Phys. Rev. Lett. {\bf 98},  044502 (2007); 
\bibitem{Berhanu07} {\sc M. Berhanu et al.}, Europhys. Lett.  {\bf 77},  59001 (2007).
\bibitem{GissingerThesis} {\sc C. Gissinger}, PhD Thesis, Pierre and Marie Curie University (2010)
\bibitem{Kono87} {\sc M. Kono}, Geoph. Res. Lett.{\bf 14}, 21-24 (1987)
\bibitem{Carbone06} {\sc V. Carbone et al.}, Phys. Rev. Lett. {\bf 96}, 128501 (2006) 
\bibitem{CK95} {\sc S.C. Cande, and D.V. Kent}, J. Geophys. Res. {\bf 100}, 6093 (1995)
\bibitem{Courtillot07} {\sc V. Courtillot and P. Olson}, Earth Planet. Sci. Lett. {\bf 260}, 495 (2007).
\bibitem{Glatzmaier99} {\sc G. Glatzmaier et al.}, Nature (London) {\bf 401}, 885 (1999)
\end{thebibliography}
\end{document}